\documentclass[twocolumn,showpacs,prb,amsmath,amssymb]{revtex4-1}
\usepackage{graphicx}
\usepackage{booktabs}
\usepackage{multirow}
\usepackage{amssymb}
\usepackage{longtable}

\begin{document}
\title{Production, Processing and Placement of Graphene and Two Dimensional Crystals}
\author{F. Bonaccorso$^1$, A. Lombardo$^1$, T. Hasan$^1$, Z. Sun$^1$, L. Colombo$^2$, A. C. Ferrari$^1$ }
\email{acf26@eng.cam.ac.uk}
\affiliation{$^1$Department of Engineering, University of Cambridge, Cambridge CB3 0FA, UK\\
$^2$ Texas Instruments Incorporated, 13121 TI Boulevard, Dallas, Texas 75243, USA}
\begin{abstract}
Graphene is at the centre of an ever growing research effort due to its unique properties, interesting for both fundamental science and applications. A key requirement for applications is the development of industrial-scale, reliable, inexpensive production processes. Here we review the state of the art of graphene preparation, production, placement and handling. Graphene is just the first of a new class of two dimensional materials, derived from layered bulk crystals. Most of the approaches used for graphene can be extended to these crystals, accelerating their journey towards applications.
\end{abstract}
\maketitle
\section*{Introduction}
Graphene has high mobility and optical transparency, in addition to flexibility, high mechanical strength and environmental stability. These properties have already had a huge impact on fundamental science\cite{Geim2007,Zhang2005,Du2009}, and are making graphene and graphene-based materials a promising platform for electronics, composites, sensors, spintronics, photonics and optoelectronics\cite{Geim2007,Charlier2008,Bonaccorso2010}. A variety of possible applications ranging from solar cells\cite{Wang2007}  and light-emitting devices\cite{Wu2009,Mayer2007} to touch screens\cite{Bae2010}, photodetectors\cite{Xia2009,Mueller2010,Echtermeyer2011,Konstantatos2012}, ultrafast lasers\cite{SunZ2010,Hasan2009}, membranes\cite{Booth2008,Chung2010}, spin valves\cite{Hill2006,Tombros2007}, high-frequency electronics\cite{Lin2010}, etc. are being explored. The present "second phase" of graphene research, after the award of the Nobel Prize to Geim and Novoselov, besides deepening the understanding of the fundamental aspects of this material, should target applications and manufacturing processes, and broaden research to other two-dimensional (2d) materials and hybrid systems. Graphene development could impact products in multiple industries, from flexible, wearable and transparent electronics, to high performance computing and spintronics. The integration of these new materials could bring a new dimension to future technologies, where faster, thinner, stronger, flexible, and broadband devices are needed\cite{Kinaret2011}. However, large-scale cost-effective production methods are required with a balance between ease of fabrication and materials quality.

\begin{figure*}
\centerline{\includegraphics[width=155mm]{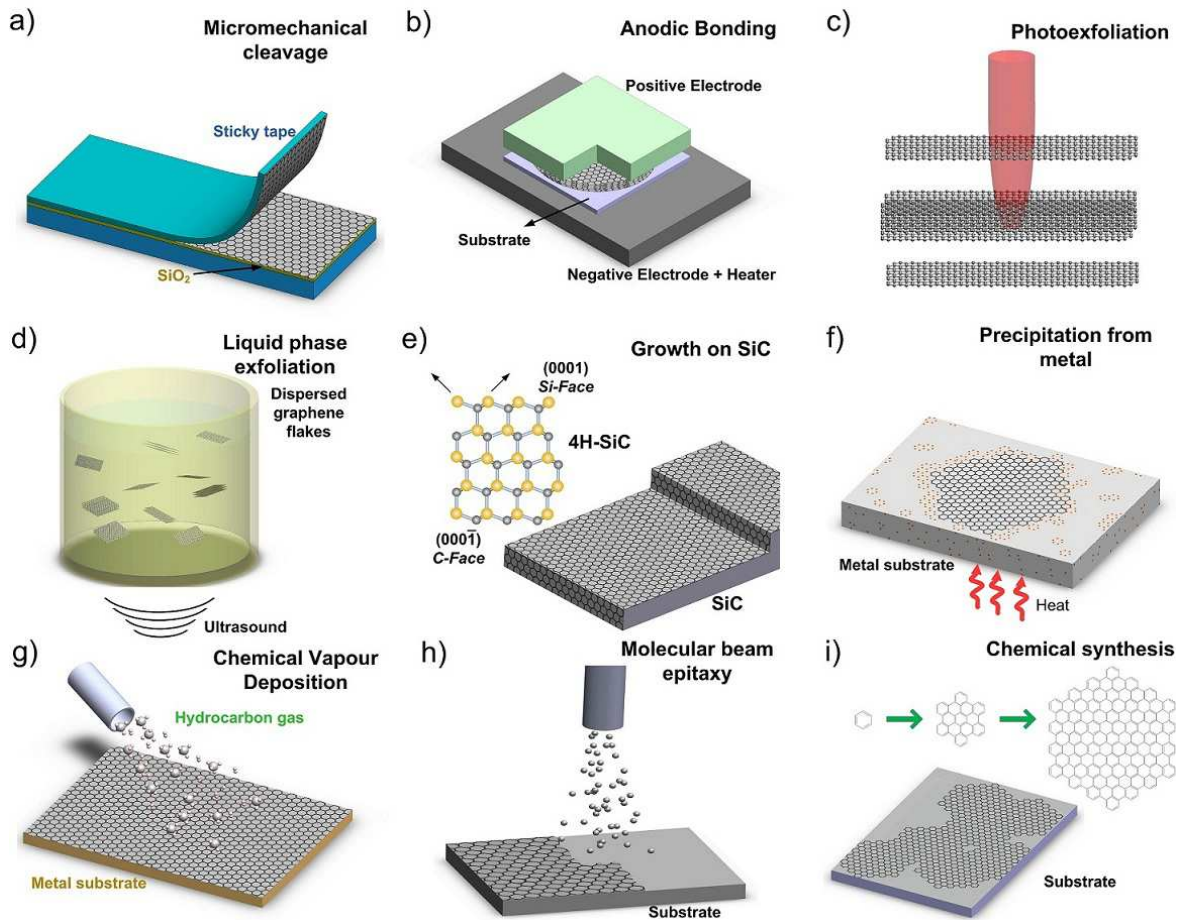}}
\caption{\label{Figure_1} Schematic illustration of the main graphene production techniques. (a) Micromechanical cleavage. (b) Anodic bonding. (c) Photoexfoliation. (d) Liquid phase exfoliation.(e) Growth on SiC. Gold and grey spheres represent Si and C atoms, respectively. At elevated T, Si atoms evaporate (arrows), leaving a carbon-rich surface that forms graphene sheets. (f) Segregation/precipitation from carbon containing metal substrate. (g) Chemical vapor deposition. (h) Molecular Beam epitaxy. (i) Chemical synthesis using benzene as building block}
\end{figure*}

Here we review the state of the art of graphene preparation, production, placement and handling, and outline how similar approaches could be used for other 2d crystals.  The main approaches are summarized in Fig.\ref{Figure_1}.

This paper is organized as follows. Section I outlines all the graphene production techniques, Section II is dedicated to processing after production, while Section III covers inorganic layered compounds and hybrid structures. Table 1 is a list of acronyms and notations.
\begin{longtable*}{llll}%
\caption {List of Acronyms}
\label{Table1}\\\hline%
1LG&Single layer graphene&$\nu$&Viscosity\\
2D&Raman 2D Peak&N&Number of graphene layers\\
2d&Two dimensional&NEMS&Nanoelectromechanical system\\
3d&Three dimensional&NLG&N-layer graphene\\
3LG&Trilayer graphene&NMP&N-MethylPyrrolidone\\
$\alpha$&Absorption coefficient&OAS&Optical absorption spectroscopy\\
a-C&Amorphous carbon&PAH&Polycyclic aromatic hydrocarbons\\
a-C:H&Hydrogenated amorphous carbon&PDMS&Poly(dimethysiloxane)\\
AFM&Atomic force microscopy&PECVD&Plasma enhanced chemical vapor deposition\\
ALD&Atomic layer deposition&PEG&Polyethylene glycol\\
ALE&Atomic layer epitaxy&PET&Poly(ethylene terephthalate)\\
BLG&Bi-layer graphene&PL&Photoluminescence\\
BMIMPF$_{6}$&1-Butyl-3-methylimidazolium hexafluorophosphate&PMMA&Poly(methyl methacrylate)\\
BN&Boron nitride&PTCDA&Perylene-3,4,9,10-tetracarboxylic dianhydride\\
c&Concentration&PV&Photovoltaic\\
CBE&Chemical beam epitaxy&PVD&Physical vapor deposition\\
CMOS&Complementary metal oxide semiconductor&QHE&Quantum Hall effect\\
CNT&Carbon nanotube&$\rho$&Density\\
CVD&Chemical vapor deposition&R2R&Roll to roll\\
DEP&Di-electrophoresis&RGO&Reduced graphene oxide\\
DGM&Density gradient medium&Rs&Sheet resistance\\
DGU&Density gradient ultracentrifugation&RT&Room temperature\\
DMF&Dimethylformamyde&RZS&Rate-zonal separation\\
DNA&Deoxyribonucleic acid&$\varrho$&Surface energy\\
FET&Field effect transistor&$\sigma$&Electrical conductivity\\
FLG&Few layer graphene&SAM&Self-assembled monolayer\\
FQHE&Fractional quantum Hall effect&SBS&Sedimentation based separation\\
$\gamma$&Surface tension&SC&Sodium cholate\\
GBL&$\gamma$-Butyrolactone&SDBS&Dodecylbenzene sulfonate\\
GIC&Graphite intercalated compounds&SDC&Sodium deoxycholate\\
GNR&Graphene nano ribbon&SLG&Single layer graphene\\
GO&Graphene oxide&STM&Scanning tunneling microscopy\\
GOIC&Graphite oxide intercalated compound&SWNT&Single wall carbon nanotube\\
GOQD&Graphene oxide quantum dots&T&Temperature\\
GQD&Graphene quantum dots&ta-C&Tetrahedral amorphous carbon\\
HBC&Hexa-perihexabenzocoronene&ta-C:H&Hydrogenated ta-C\\
$\eta$&Carrier mobility&ta-C:N&Nitrogenated ta-C\\
\textit{h}-BN&Hexagonal boron nitride&TCF&Transparent conducting film\\
hcp&Hexagonal closed packed&TEM&Transmission electron microscopy\\
HMIH&1-hexyl-3-methylimidazolium hexafluorophosphate&TGA&Thermo-gravimetric analysis\\
ICP&Inductively coupled plasma&TLG&Tri-layer graphene\\
IL&Ionic liquid&TMD&Transition metal dichalcogenide\\
LEED&Low-energy electron diffraction&TMO&Transition metal oxide\\
LM&Layered material&UHV&Ultra high vacuum\\
LPCVD&Low pressure chemical vapor deposition&UV&Ultra violet\\
LPE&Liquid phase exfoliation&VRH&Variable range hopping\\
m&Staging index&XPS&X-ray photoelectron spectroscopy\\
MBE&Molecular beam epitaxy&Y$_M$&Yield by SLG percentage\\
MC&Micromechanical cleavage&Y$_W$&Yield by weight\\
MLG&Multilayer graphene&Y$_{WM}$&Yield by SLG weight\\
\\\hline
\end{longtable*}

\begin{figure}
 \centerline{\includegraphics[width=80mm]{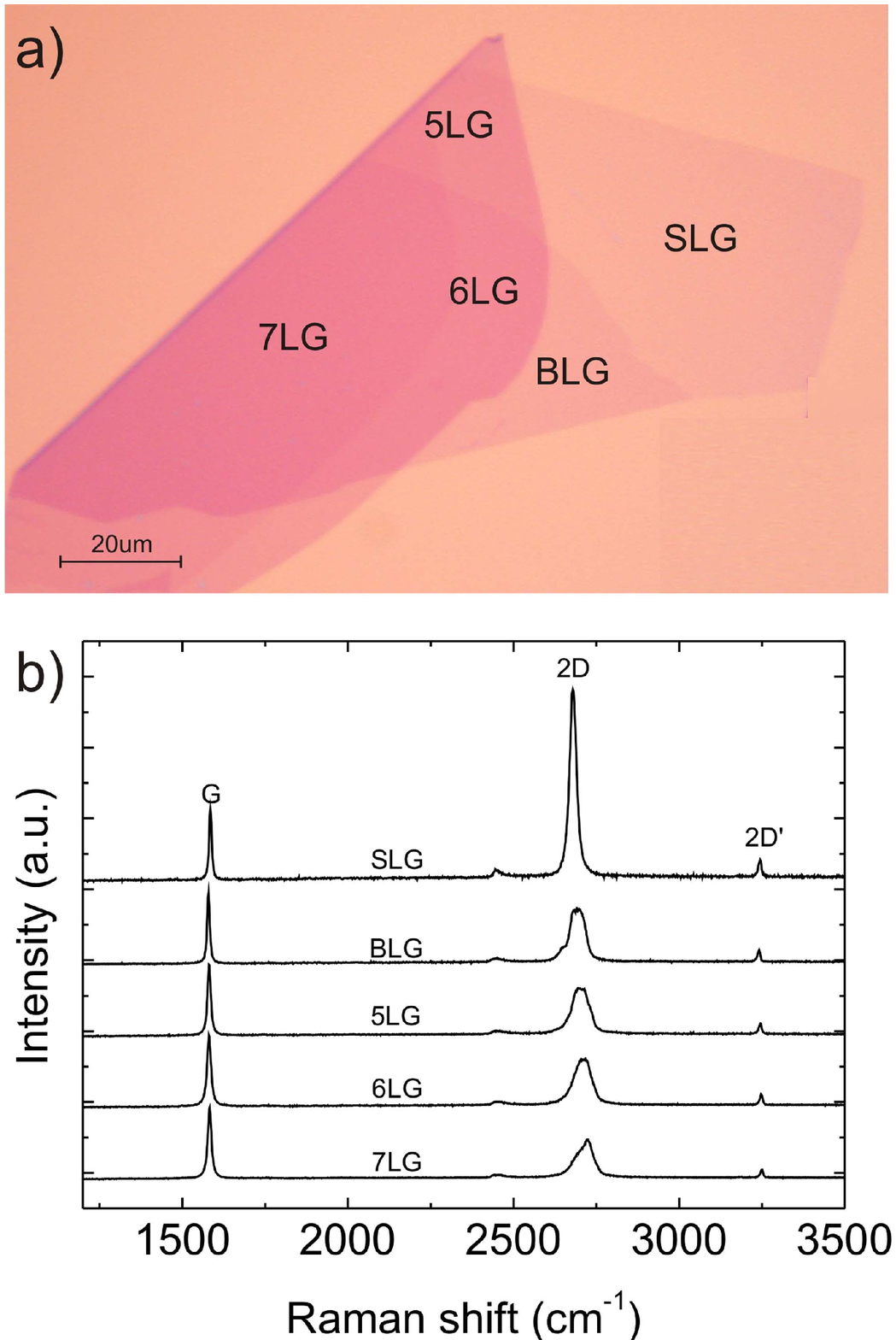}}
 \caption{\label{Figure_2}(a) Optical micrograph of MC flake, consisting of regions of different thickness. (b) Evolution of Raman spectra with number of layers\cite{Ferrari2006}. The spectra are normalized to have the same G peak intensity.}
\end{figure}
\section{\label{Pro}Graphene production}
\subsection{\label{DRE}Dry exfoliation}
Dry exfoliation is the splitting of layered materials (LM) into atomically thin sheets via mechanical, electrostatic, or electromagnetic forces in air, vacuum or inert environments.
\subsubsection{\label{MC}Micromechanical cleavage}
Micromechanical cleavage (MC), also known as micromechanical exfoliation, has been used for decades by crystal growers and crystallographers\cite{Schultz1994,Romero1992}. In 1999 Ref.[\onlinecite{Lu1999}] reported a controlled method of cleaving graphite, yielding films consisting of several layers of graphene. Ref.[\onlinecite{Lu1999}] also suggested that "more extensive rubbing of the graphite surface against other flat surfaces might be a way to get multiple or even single atomic layers of graphite plates." This was then firstly demonstrated, achieving SLG using an adhesive tape, by Novoselov \textit{et al.} [\onlinecite{Novoselov2005}], as illustrated in Fig.\ref{Figure_1}a.

Micromechanical cleavage is now optimized to yield high quality layers, with size limited by the single crystal grains in the starting graphite, of the order of millimeters\cite{Geim2009}. The number of layers can be readily identified by elastic\cite{Casiraghi2007a} and inelastic\cite{Ferrari2006} light scattering. Raman spectroscopy also allows a fast and non-destructive monitoring of doping\cite{Casiraghi2007APL,Pisana2007,Das2008}, defects\cite{Ferrari2000,Ferrari2001,Ferrari2004,Cancado2011}, strain\cite{Ferralis2008,Mohiuddin2009}, disorder\cite{Ferrari2007}, chemical modifications\cite{Ferrari2001,Elias2009} and edges\cite{Casiraghi2009,Basko2009}, see Fig.\ref{Figure_2}.  Mobilities of up to 10$^7$ cm$^2$ V$^{-1}$ s$^{-1}$ at 25K were reported for a decoupled single layer graphene (SLG) on the surface of bulk graphite\cite{Neugebauer2009}, and up to 10$^6$ cm$^2$ V$^{-1}$ s$^{-1}$ on current-annealed suspended SLGs\cite{Elias2011}, while room temperature (RT) mobilities up to$\sim$20,000cm$^2$ V$^{-1}$ were measured in as-prepared SLGs\cite{Ni2010}. Suspended SLGs, cleaned by current annealing (see Sect.\ref{Cleaning}), can reach mobilities of several 10$^6$cm$^2$ V$^{-1}$ s$^{-1}$ [\onlinecite{Mayorov2012}]. Mobilities in excess of 10$^5$cm$^2$ V$^{-1}$ s$^{-1}$, with ballistic transport at the micron level, were reported for SLG encapsulated between exfoliated hexagonal boron nitride (h-BN) layers\cite{Mayorov2011a}.

Although MC is impractical for large scale applications, it is still the method of choice for fundamental studies. Indeed, the vast majority of basic results and prototype devices were obtained using MC flakes. Thus, MC remains ideal to investigate both new physics and new device concepts.
\subsubsection{\label{AB}Anodic bonding}
Anodic bonding is widely used in the microelectronics industry to bond Si wafers to glass\cite{Albaugh1991}, to protect them from humidity or contaminations\cite{Henmi1994}. When employing this technique to produce SLGs\cite{Shukla2009,Moldt2011}, graphite is first pressed onto a glass substrate, and a high voltage of few KVs (0.5-2 kV) is applied between the graphite and a metal back contact (see Fig.\ref{Figure_1}b), and the glass substrate is then heated ($\sim$200$^\circ$C for$\sim$10-20mins)\cite{Shukla2009,Moldt2011}. If a positive voltage is applied to the top contact, a negative charge accumulates in the glass side facing the positive electrode, causing the decomposition of Na$_2$O impurities in the glass into Na$^+$ and O$_{2}^{-}$ ions\cite{Shukla2009,Moldt2011}. Na$^+$ moves towards the back contact, while O$_{2}^{-}$ remains at the graphite-glass interface, establishing a high electric field at the interface. A few layers of graphite, including SLGs, stick to the glass by electrostatic interaction and can then be cleaved off\cite{Shukla2009,Moldt2011}; temperature and applied voltage can be used to control the number of layers and their size\cite{Shukla2009,Moldt2011}. Anodic bonding has been reported to produce flakes up to about a millimeter in width\cite{Shukla2009}.
\subsubsection{\label{LA}Laser ablation and photoexfoliation}
Laser ablation is the use of a laser beam to remove material from a solid surface\cite{Douglas2003}. If the irradiation results into the detachment of an entire or partial layer, the process is called photoexfoliation\cite{Miyamoto2010}.

Laser pulses can in principle be used to ablate/exfoliate graphite flakes, Fig.\ref{Figure_1}(c). Indeed, tuning the energy density permits the accurate patterning of graphene\cite{Dhar2011}. The ablation of a defined number of layers can be obtained exploiting the energy density windows required for ablating a SLG\cite{Dhar2011} and N-layer graphene (NLGs) of increasing number of layers\cite{Dhar2011}. Ref.[\onlinecite{Dhar2011}] reported that energy density increases for decreasing N up to$\sim$7LG. Ref.[\onlinecite{Dhar2011}] argued that the N dependence of the energy density is related to the coupling of heat with NLGs via phonons, with the specific heat scaling as 1/N. For N$>$7 the ablation threshold saturates\cite{Dhar2011}.

Laser ablation is still in its infancy\cite{Dhar2011,Reininghaus2012}, and needs further development. The process is best implemented in inert or vacuum conditions\cite{Lee2010,Qian2011} since ablation in air tends to oxidize the graphene layers\cite{Dhar2011}. Promising results were recently demonstrated also in liquids\cite{Mortazavi2012}.
\subsection{\label{LPE}Liquid-Phase-Exfoliation (LPE)}
Graphite can also be exfoliated in liquid environments exploiting ultrasounds to extract individual layers, Fig.\ref{Figure_1}d. The liquid-phase exfoliation (LPE) process generally involves three steps: 1) dispersion of graphite in a solvent; 2) exfoliation; 3) "purification". The third step is necessary to separate exfoliated from un-exfoliated flakes, and is usually carried out via ultracentrifugation.

The LPE yield can be defined in different ways. The yield by weight, Y$_W$ [\%], is defined as the ratio between the weight of dispersed graphitic material and that of the starting graphite flakes\cite{Hernandez2008}. The yield by SLG percentage, Y$_M$ [\%], is defined as the ratio between the number of SLG and the total number of graphitic flakes in the dispersion\cite{Hernandez2008}. The Yield by SLG weight, Y$_{WM}$ [\%], is defined as the ratio between the total mass of dispersed SLG and the total mass of all dispersed flakes. Y$_W$ does not give information on the the amount of SLG, but only the total amount of graphitic material. Y$_M$ [\%], Y$_{WM}$ [\%] are more suitable to quantify the dispersed SLGs.

In order to determine Y$_W$ it is necessary to calculate the concentration c [g L$^{-1}$] of dispersed graphitic material. c is usually determined via optical absorption spectroscopy (OAS)\cite{Hernandez2008,Lotya2009,Lotya2010,Khan2010,Hasan2010,Torrisi2012}, exploiting the Beer-Lambert Law: A=$\alpha$cl, where l [m] is the length of the optical path and $\alpha$ [L g$^{-1}$ m$^{-1}$] is the absorption coefficient. $\alpha$ can be experimentally determined by filtering a known volume of dispersion, \textit{e.g.} via vacuum filtration, onto a filter of known mass\cite{Hernandez2008,Lotya2009,Lotya2010,Khan2010}, and measuring the resulting mass using a microbalance. The filtered material is made up of a graphitic mass, surfactant or solvents and residual from the filter\cite{Hernandez2008,Lotya2009}. Thermogravimetric (TGA) analysis is used to determine the weight percentage of graphitic material in it, thus enabling the measurement of c\cite{Hernandez2008,Lotya2009,Lotya2010,Khan2010}. However, different values of $\alpha$ have been estimated both for aqueous\cite{Lotya2009,Lotya2010} and non-aqueous dispersions\cite{Hernandez2008,Khan2010}. Ref.[\onlinecite{Hernandez2008}] derived $\alpha\sim$2460mLmg$^{-1}$m$^{-1}$ for a variety of solvents, \textit{i.e.} N-MethylPyrrolidone, NMP, Dimethylformamyde, DMF, Benzyl benzoate,  $\gamma$-Butyrolactone, GBL, etc., while later Ref.[\onlinecite{Khan2010}] reported  $\alpha\sim$3620mL mg$^{-1}$ m$^{-1}$ for NMP. Ref.[\onlinecite{Lotya2009}] gave $\alpha\sim$1390mL mg$^{-1}$ m$^{-1}$ for aqueous dispersions with sodium
dodecylbenzene sulfonate (SDBS), while Ref.[\onlinecite{Lotya2010}] reported$\sim$6600mL mg$^{-1}$ m$^{-1}$, still for aqueous dispersions but with sodium cholate (SC). Ref.[\onlinecite{Lotya2010}] assigned this discrepancy to the c difference between the two dispersions. However, $\alpha$ cannot be dependent on c (indeed it is used for its determination), thus more work is needed to determine its exact value.

Y$_M$ is usually determined via transmission electron microscopy (TEM) and atomic force microscopy (AFM). In TEM, N can be counted both analyzing the edges\cite{Ferrari2006} of the flakes and by using electron diffraction patterns\cite{Ferrari2006}. AFM enables the estimation of N by measuring the height of the deposited flakes and dividing by the graphite interlayer distance. However, the estimation for the height of SLG via AFM is dependent on the substrate. Indeed, for SiO$_2$ a SLG has an height of $\sim$1nm\cite{Novoselov2005}, while on mica is $\sim$0.4nm\cite{Valles2008}. Raman spectroscopy is used for the determination of Y$_M$\cite{Hernandez2008,Hasan2010,Torrisi2012} and to confirm the results obtained with TEM and/or AFM.

Y$_{WM}$ [\%] requires the estimation of SLGs area other than N\cite{Hernandez2008}. However, although this is a more accurate parameter (giving quantitative and qualitative information on SLGs), with respect Y$_W$ and Y$_M$, to characterize a dispersion, its determination is very time consuming. Indeed, to the best of our knowledge it was used only once, when it was defined\cite{Hernandez2008}. However, for a semi-quantitative evaluation of the dispersion Y$_M$ and Y$_W$ must be reported if Y$_{WM}$ is not.
\subsubsection{\label{LPEG}LPE of graphite}
\begin{figure*}
\centerline{\includegraphics[width=160mm]{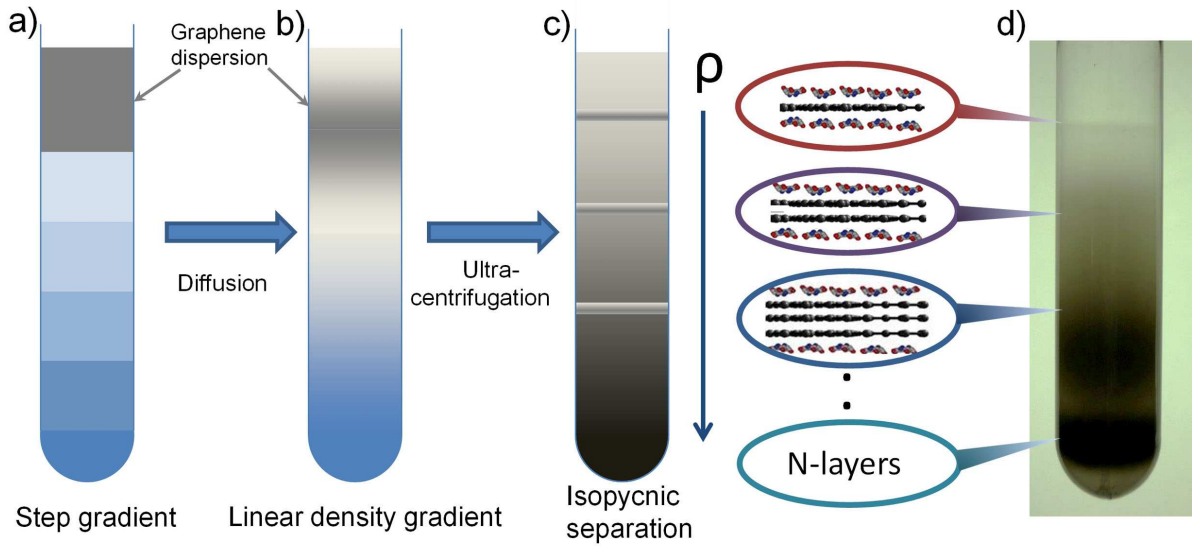}}
\caption{\label{Figure_3} Sorting of graphite flakes via isopycnic separation. (a) Formation of step gradient by placing a density gradient medium with decreasing concentration. (b) Linear density gradient formed via diffusion. (c) During isopycnic separation the graphite flake-surfactant complexes move along the cuvette, dragged by the centrifugal force, until their isopycnic points. The buoyant density of the flake-surfactant complexes increases with number of layers. (d) Photograph of cuvette containing sorted flakes.}
\end{figure*}
Graphene flakes can be produced by exfoliation of graphite via chemical wet dispersion followed by ultrasonication in water\cite{Lotya2009,Green2009,Marago2010,Hasan2010} and organic solvents\cite{Hasan2010,Hernandez2008,Blake2008,Torrisi2012}. Ultrasound-assisted exfoliation is controlled by hydrodynamic shear-forces, associated with cavitation\cite{Mason1999}, \textit{i.e.} the formation, growth, and collapse of bubbles or voids in liquids due to pressure fluctuations\cite{Mason1999}. After exfoliation, the solvent-graphene interaction needs to balance the inter-sheet attractive forces.

Solvents ideal to disperse graphene are those that minimize the interfacial tension [mN/m] between the liquid and graphene flakes (i.e. the force that minimizes the area of the surfaces in contact)\cite{Israelachvili} . In general, interfacial tension plays a key role when a solid surface is immersed in a liquid medium\cite{Israelachvili,Lyklema1999,Ghatee2005}. If the interfacial tension between solid and liquid is high, there is poor dispersibility of the solid in the liquid\cite{Israelachvili}. In the case of graphitic flakes in solution, if the interfacial tension is high, the flakes tend to adhere to each other and the work of cohesion between them is high (i.e. the energy per unit area required to separate two flat surfaces from contact\cite{Israelachvili}), hindering their dispersion in liquid. Liquids with surface tension (\textit{i.e.} the property of the surface of a liquid that allows it to resist an external force, due to the cohesive nature of its molecules\cite{Israelachvili}) $\gamma$$\sim$40mN/m [\onlinecite{Hernandez2008}], are the "best" solvents for the dispersion of graphene, since they minimize the interfacial tension between solvent and graphene.

Ref.[\onlinecite{WangLang}] determined via wettability and contact angle measurements the surface energy, $\varrho$ [mJ/m$^2$],  of different graphitic materials, finding $\varrho\sim$46mJ/m$^2$,$\sim$55mJ/m$^2$,$\sim$62mJ/m$^2$ for reduced graphene oxide (RGO), graphite and graphene oxide (GO). The slight difference being due to the different surface structure of GO, RGO and graphite. Ref.[\onlinecite{Shin2010}] reported that contact angle measurements are not affected by N.

The majority of solvents with $\gamma\sim$40mN/m (\textit{i.e.} NMP, DMF, Benzyl benzoate, GBL, etc.) [see Ref.[\onlinecite{Hernandez2008}] for a complete list] have some disadvantages. E.g., NMP may be toxic for the reproductive organs\cite{Solomon1995}, while DMF may have toxic effects on multiple organs\cite{Kennedy1986}. Moreover, all have high ($>$450K) boiling points, making it difficult to remove the solvent after exfoliation. As an alternative, low boiling point solvents\cite{ONeill2011}, such as acetone, chloroform, isopropanol, etc. can be used. Water, the "natural" solvent, has $\gamma\sim$72mN/m [\onlinecite{Israelachvili}], too high (30mN/m higher than NMP) for the dispersion of graphene\cite{WangLang} and graphite\cite{WangLang}. In this case, the exfoliated flakes can be stabilized against re-aggregation by Coulomb repulsion using linear chain surfactants, \textit{e.g.} SDBS\cite{Lotya2009}, or bile salts \textit{e.g.} SC\cite{Green2009} and sodium deoxycholate (SDC)\cite{Marago2010,Hasan2010}, or polymers \textit{e.g.} pluronic\cite{Seo2011}, etc. However, depending on the final application, the presence of surfactants/polymers may be an issue, e.g. compromising, decreasing, the inter-flake conductivity\cite{Nirmalraj2011}.

Thick flakes can be removed by different strategies based on ultracentrifugation in a uniform medium\cite{Svedberg1940}, or in a density gradient medium (DGM)\cite{Behrens1939}. The first is called differential ultracentrifugation (sedimentation based-separation, SBS)\cite{Svedberg1940}, while the second is called density gradient ultracentrifugation (DGU)\cite{Behrens1939}. The SBS process separates various particles on the basis of their sedimentation rate\cite{Svedberg1940} in response to a centrifugal force acting on them. Sedimentation based separation is the most common separation strategy and, to date, flakes ranging from few nanometers to a few microns have been produced, with concentrations up to a few mg/ml\cite{Khan2010, Alzari2011}. High concentration is desirable for large scale production of composites\cite{Hernandez2008} and inks\cite{Torrisi2012}. Y$_M$ up to $\sim$70\%  were achieved by mild sonication in water with SDC, followed by SBS\cite{Marago2010}, while Y$_M\sim$33\% was reported with NMP\cite{Torrisi2012}. This Y$_M$ difference is related to the difference in flake lateral size. In water-surfactant dispersions flakes are on average smaller ($\sim$30nm\cite{Marago2010} to$\sim$200nm\cite{Lotya2009}) than in NMP($\sim$1$\mu$m\cite{Hernandez2008,Torrisi2012}), since the viscosity ($\nu$) at RT of NMP (1.7mPas\cite{Lide2005}) is higher than water ($\sim$1mPas\cite{Lide2005}). Larger flakes in a higher viscosity medium experience a higher frictional force\cite{Svedberg1940,Behrens1939} that reduces their sedimentation coefficient, making it more difficult for them to sediment. This decreases Y$_M$ in NMP compared to water.

During DGU, the flakes are ultracentrifuged in a preformed DGM\cite{Behrens1939,Williams1958}, see Figs.\ref{Figure_3}a,b, where they move along the cuvette until they reach the corresponding isopycnic point, \textit{i.e.}, the point where their buoyant density equals that of the surrounding DGM\cite{Behrens1939}. The buoyant density is defined as the density ($\rho$) of the medium at the corresponding isopycnic point\cite{Behrens1939,Williams1958}. Isopycnic separation was used to sort nanotubes by diameter\cite{Arnold2005,Crochet2007}, metallic vs semiconducting nature\cite{Arnold2006}  and chirality\cite{BonaDGU}. However, unlike nanotubes of different diameter, graphitic flakes have the same density, irrespective of N, so another approach is needed to induce a density difference: coverage of the flakes with a surfactant results in an increase of buoyant density with N, Fig.\ref{Figure_3}c. Fig.\ref{Figure_3}d shows a cuvette after isopycnic separation. Ref.[\onlinecite{Green2009}] reported Y$_M\sim$80\% for this technique with SC surfactant.

Another method is the so-called rate zonal separation (RZS)\cite{SunX2010}. This exploits the difference in sedimentation rates of nanoparticles with different size\cite{Tyler2011}, shape\cite{Akbulut2012} and mass\cite{Tyler2011}, instead of the difference in nanoparticle density, as in the case of isopycnic separation. RZS was used to separate flakes with different size\cite{SunX2010} (the larger the size, the larger the sedimentation rate).

Other routes based on wet chemical dispersion have been investigated, such as exfoliation in ionic liquids (ILs)\cite{Nuvoli2011,WangChem2010}, 1-hexyl-3-methylimidazolium hexafluorophosphate (HMIH)\cite{Nuvoli2011} or 1-butyl-3-methylimidazolium bis(trifluoro-methane-sulfonyl)imide ([Bmim]-[Tf2N])\cite{WangChem2010}. These are a class of purely ionic, salt-like materials\cite{Welton1999}, defined as salts in the liquid state (below 100$^{\circ}$C), largely made of ions\cite{Welton1999}. Ref.[\onlinecite{Nuvoli2011}] reported concentrations exceeding 5mg/mL by grinding graphite in a mortar with ILs, followed by ultrasonication and centrifugation. The flakes had sizes up to$\sim$3-4$\mu$m, however no data was shown for N\cite{Nuvoli2011}. Ref.[\onlinecite{Nuvoli2011}] used a long ultrasonication process ($>$24 hours), probably because of the IL high viscosity. In SBS viscosity plays a fundamental role. Flakes in a higher viscosity medium have a lower sedimentation coefficient with respect to water. The sedimentation coefficient is commonly measured in Svedberg (S) units (with 1S corresponding to 10$^{-13}$sec.), the time needed for particles to sediment out of the fluid, under a centrifugal force\cite{Svedberg1940}. E.g., for a flake dispersed in [Bmim]-[Tf$_2$N]  ($\rho$=1.43g/cm$^3$, $\nu$=32mPas), the sedimentation coefficient is$\sim$55 times smaller than in water. There are no reports to date showing that exfoliation via ultrasonication in ILs can have the same Y$_M$ as in water\cite{Marago2010}, or organic solvents\cite{Torrisi2012}. Moreover, the resultant flakes contain oxygen functional groups\cite{WangChem2010}, probably due strong non-covalent interactions, or covalent functionalization with [Bmim][Tf$_{2}$N] itself\cite{WangChem2010}. A step forward for the production of flakes without these functional groups was reported in Ref.[\onlinecite{Shang2012}], where oxygen-free flakes were made by grinding graphite in 1-Butyl-3-methylimidazolium hexafluorophosphate, [BMIMPF$_{6}$]. Ionic liquids were then removed by mixing with Acetone and DMF\cite{WangChem2010}. Controlling grinding time and IL quantity, Ref.[\onlinecite{WangChem2010}] reported graphitic quantum dots (GQDs) with size from 9 to 20nm and thickness between 1 and 5nm.

An alternative process is non-covalent functionalization with 1-pyrenecarboxylic acid, as reported in Ref.\onlinecite{An2010}. However, Ref.[\onlinecite{An2010}] only achieved a mixture of SLGs and FLGs. Thus, work is still needed to improve Y$_M$.
\begin{figure}
 \centerline{\includegraphics[width=90mm]{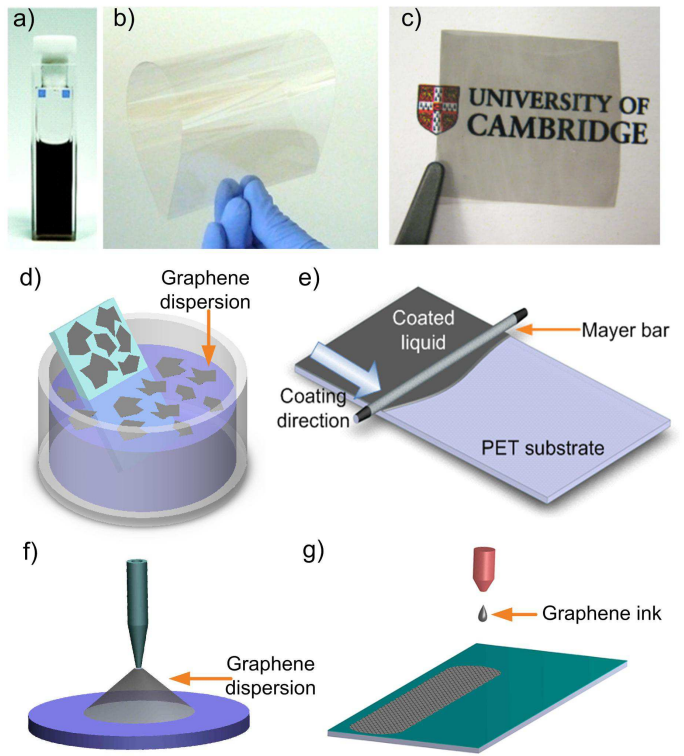}}
 \caption{\label{Figure_4}(a) Graphene ink produced via LPE of graphite\cite{Torrisi2012}. (b) Graphene-based transparent and flexible conductive film. (c) Graphene polymer composite produced via LPE of graphite in water and mixed with Polyvinyl alcohol\cite{Hasan2009,Hasan2010}. (d) Dip casting of LPE graphene. The substrate is immersed in the graphene dispersion/ink to obtain a uniform coverage\cite{Wang2007}. (e) Rod coating of LPE graphene. In this coating process, a wire-covered metal bar (Meyer bar) is used to apply in a controlled way the graphene dispersion onto the substrate. (f) Spray coating. The graphene dispersion/ink is deposited through the air onto the substrate by a spray (i.e. spray gun)\cite{Blake2008}. (g) Ink-jet printing is used to deposit droplets of graphene inks\cite{Torrisi2012} on substrates with higher precision with respect to other approaches, such as dip casting, rod and spray coating.}
\end{figure}

LPE is cheap and easily scalable, and does not require expensive growth substrates. Furthermore it is an ideal means to produce inks\cite{Torrisi2012} (Fig.\ref{Figure_4}a), thin films\cite{Hernandez2008} (Fig.\ref{Figure_4}b), and composites\cite{Hasan2009, Hasan2010} (Fig.\ref{Figure_4}c). The resulting material can be deposited on different substrates (rigid and flexible) by drop and dip casting\cite{Wang2007} (Fig.\ref{Figure_4}d), rod coating (Fig.\ref{Figure_4}e), spray coating\cite{Blake2008} (Fig.\ref{Figure_4}f), screen and ink-jet printing\cite{Torrisi2012} (Fig.\ref{Figure_4}g), vacuum filtration\cite{Hernandez2008}, Langmuir-Blodgett\cite{Li2008W}, and other techniques discussed in Sect.\ref{inks}.

LPE flakes have limited size due to both the exfoliation procedure, that induces in-plane fracture, and the purification process, which separates large un-exfoliated flakes. To date, LPE-SLGs have area mostly below 1$\mu$m$^2$ [Refs.\onlinecite{Lotya2009,Green2009,Marago2010,Hasan2010,Hernandez2008,Torrisi2012,ONeill2011,Khan2010}].

Liquid phase exfoliation can also be optimized to produce graphene nanoribbons (GNRs), with widths $<$10nm\cite{Li2008}. Ref.[\onlinecite{Li2008}] ultrasonicated expanded graphite\cite{Zheng2003}, \textit{i.e.} with larger interlayer distance with respect to graphite due to intercalation of nitric\cite{Forsman1978} and sulfuric acid\cite{McAllister2007}. Expanded graphite was dispersed in a 1,2-dichloroethane solution of poly(m-phenylenevinylene-co- 2,5-dioctoxy-p-phenylenevinylene), ultrasonicated and ultracentrifuged, resulting in a combination of flakes and GNRs of different shapes. However, the GNR production mechanism via LPE of graphite is not well understood. Thus, more work is needed to fully understand and improve GNRs production via LPE.
\subsubsection{\label{LPEGO}LPE of graphite oxide}
LPE is a versatile technique and can be exploited not only for the exfoliation of pristine graphite as reported in Sect.\ref{LPEG} but also for the exfoliation of graphite oxide and graphite intercalated compounds (GICs), which have different properties with respect to pristine graphite. The oxidation of graphite in the presence of potassium chlorate (KClO$_3$) and fuming nitric acid was developed by Brodie in 1859 while investigating the reactivity of graphite flakes\cite{Brodie1860}. This process involved successive oxidative treatments of graphite in different reactors\cite{Brodie1860}. In 1898, Staudenmaier modified Brodie's process by using concentrated sulphuric acid and adding KClO$_3$ in successive steps during the reaction\cite{Staudenmaier1898}. This allowed carrying out the reaction in a single vessel, streamlining the production process\cite{Dreyer2010}. However, both methods were time consuming and hazardous, as they also yielded chlorine dioxide (ClO$_2$) gas\cite{Hyde1904}, which can explosively decompose into oxygen and chlorine\cite{Hyde1904}. Graphite oxide flakes were already investigated by Kohlschütter and Haenni in 1918\cite{Kohlschutter1918}, and the first TEM images reported in 1948 by Ruess and Vogt\cite{Ruess1948} showed the presence of single GO sheets. In 1958, Hummers modified the process using a mixture of sulphuric acid, sodium nitrate and potassium permanganate\cite{Hummers1958}. Avoiding KClO$_3$ made the process safer, quicker, with no explosive byproducts\cite{Hummers1958}.
\begin{figure*}
\centerline{\includegraphics[width=160mm]{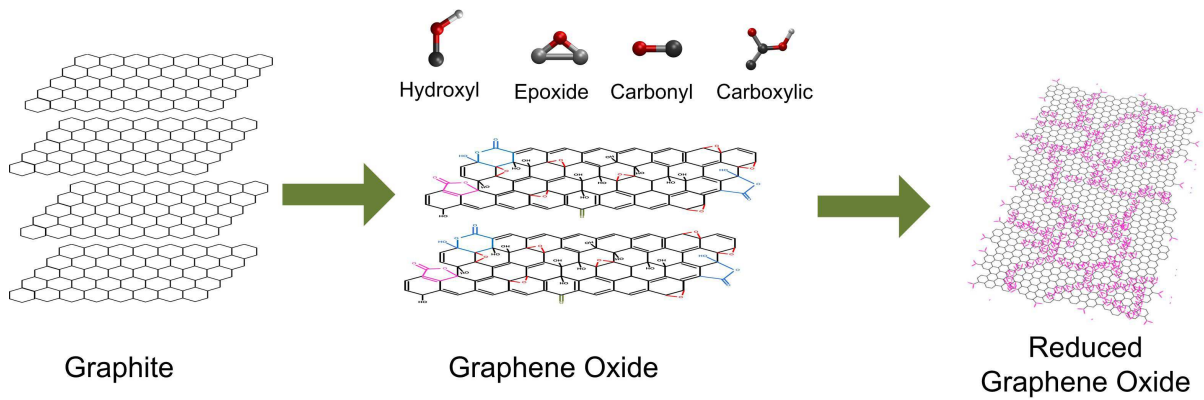}}
\caption{\label{Figure_5} GO synthesis and reduction. Graphite can be oxidized with different procedures in the presence of strong acids. The GO flakes have the plane functionalized with epoxy and hydroxyl groups both above and below and at the edges\cite{Paredes2008,Eda2009}. A partial recovery of the electronic properties can be reached following a reduction treatment\cite{Mattevi2009,Li2008,Niyogi2006,Schniepp2006,Lomeda2008,Paredes2008}. However, none of the current approaches can fully remove the defects.}
\end{figure*}

These aggressive chemical processes disrupt the sp$^2$-bonded network and introduce hydroxyl or epoxide groups\cite{Mattevi2009,Cai2008,Boehm} in the basal plane, while carbonyl and carboxylic groups, together with lactone, phenol and quinone attach to the edges (see Fig.5). However, the introduction of these functional groups is essential for the GO production and subsequent liquid dispersion.

GO flakes can be prepared via sonication\cite{Li2008,Stankovich2007}, stirring\cite{Lomeda2008}, thermal expansion\cite{Schniepp2006}, etc., of graphite oxide. The aforementioned functional groups make GO flakes strongly hydrophilic, allowing their dispersion in pure water\cite{Li2008,Stankovich2007}, organic solvents\cite{Niyogi2006,Schniepp2006,Lomeda2008}, aqueous mixtures with methanol, acetone, acetonitrile\cite{Paredes2008}  or 1-propanol and ethylene glycol\cite{Si2008}. However, although large GO flakes, up to several microns\cite{Su2009a}, can be produced, they are defective\cite{Mattevi2009} and insulating, with sheet resistance (Rs)$\sim$10$^{12}\Omega/\Box$, or higher\cite{Becerril2008}.

GO is luminescent under continuous wave irradiation\cite{Gokus2009}. Visible excitation gives a broad photoluminescence (PL) spectrum from visible to near-infrared\cite{SunX2008}, while blue emission\cite{Eda2009} is detected upon ultraviolet (UV) excitation. This makes GO an interesting material for lighting applications (e.g. light emitting devices\cite{Matyba2011}) and bio-imaging\cite{SunX2008}.

Several processes have been developed to chemically "reduce" the GO flakes, \textit{i.e.} decrease the oxidation state of the oxygen-containing groups in order to re-establish an electrical and thermal conductivity as close as possible to pristine graphene. In 1962, the reduction of graphite oxide in alkaline dispersions was proposed for the production of thin (down to single layer) graphite lamellaes\cite{Boehm,Boehm1962}. Other methods involve treatments by hydrazine\cite{Li2008,Gomez2007}, hydrides\cite{Si2008,Bourlinos2003}, p-phynylene\cite{Chen2009}, hydroquinone\cite{Bourlinos2003} etc, as well as dehydration\cite{Liao2011} or thermal reduction\cite{Mattevi2009,Schniepp2006,Wang2009}. UV-assisted photocatalyst reduction of GO was also proposed\cite{Williams2008}, whereby GO reduces as it accepts electrons from UV irradiated TiO$_2$ nanoparticles\cite{Williams2008}.

The charge transport in RGO is believed to take place via variable-range hopping (VRH)\cite{Eda2009,Kaiser2009}. Individual RGO sheets have been prepared with electrical conductivity ($\sigma)\sim$350Scm$^{-1}$ [\onlinecite{Lopez2009}], while higher values (1314Scm$^{-1}$) were achieved in thin films\cite{Su2009}, because in the latter RGO flakes are equivalent to resistors in parallel\cite{Gomez2007}. These $\sigma$ are much bigger than organic semiconductors (e.g. poly($\beta$'-dodecyloxy(-$\alpha$,$\alpha$'-$\alpha$',$\alpha$"-)terthienyl) (poly(DOT)) $\sim$10$^{-3}$Scm$^{-1}$ for a sample doped to$\sim$10$^{21}$cm$^{-3}$)\cite{Brown1997}.

It is important to differentiate between dispersion-processed flakes, retaining the graphene electronic properties, such as those reported in Refs.[\onlinecite{Lotya2009,Green2009,Marago2010,Hasan2010,Hernandez2008,Torrisi2012,Khan2010,Blake2008}], and GO flakes, such as those in Refs.[\onlinecite{Li2008,Stankovich2007,Niyogi2006,Schniepp2006,Lomeda2008}]. Indeed, GO can have $\sigma$ as low as$\sim$10$^{-5}$Scm$^{-1}$[\onlinecite{Li2008}], while LPE graphene can feature $\sigma$ up to$\sim$10$^4$Scm$^{-1}$ [\onlinecite{Blake2008}].

GO and RGO can be deposited on different substrates with the same techniques used for LPE graphene, discussed in Sect.[\ref{inks}]. GO and RGO are ideal for composites\cite{Stankovich2006}, due the presence of functional groups, which can link polymers\cite{Stankovich2006}.

Ref.[\onlinecite{Korkus2011}] reported RGO sheets with $\sigma\sim$10$^3$Sm$^{-1}$, high flexibility, and surface areas comparable to SLG, thus interesting for a range of electronic and optoelectronic applications. Thin films of RGO have been tested as field-effect transistors (FETs)\cite{Lin2008}, transparent conducting films (TCFs)\cite{Eda2008}, electro-active layers\cite{Mativetsky2010}, solar cells\cite{Wang2008}, ultrafast lasers\cite{Sun2011C,Bao2010}, etc. Patterning has been used to create conductive RGO-based electrodes\cite{Eda2009}.
\begin{figure}
\centerline{\includegraphics[width=85mm]{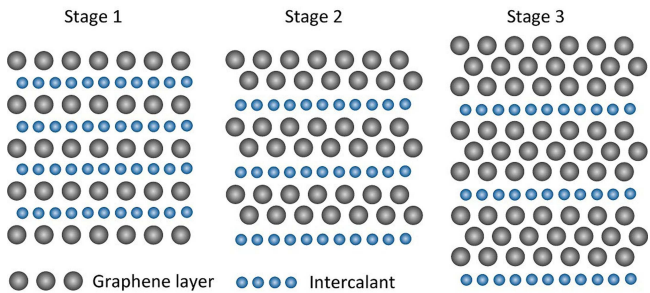}}
\caption{\label{Figure_6} Graphite intercalation compounds. In stage 1, SLG alternate with intercalant layers. In stage 2, stage 3, etc, 2, 3, etc. graphene layers separate two intercalant layers}
\end{figure}
\subsubsection{\label{LPEGIC}LPE of intercalated graphite}
GIC are formed by periodic insertion of atomic or molecular species (intercalants) between graphite layers\cite{Dresselhaus2002,Inagaki1989,Boehm1994}. GICs are classified in terms of "staging" index m, \textit{i.e.} the number of graphene layers between two intercalant layers. Thus, e.g., a stage 3 GIC (see Fig.\ref{Figure_6}) has each 3 adjacent graphene layers sandwiched by 2 intercalant layers\cite{Dresselhaus2002} (the latter can also be more than 1 atom thick).

Ref.[\onlinecite{Dresselhaus2002,Inagaki1989}] summarized the historical development of GICs. The production of GICs started in the mid-1800s with the seminal work of Schaffautl in 1841\cite{Schaffautl1841}. The first m determination by X-Ray diffraction was done in 1931 by Hoffman and Fenzel\cite{Hoffman1931}. Systematic studies started in the late 1970s.

Intercalation of atoms or molecules with different m gives rise to a wide variety of electrical\cite{Dresselhaus2002}, thermal\cite{Dresselhaus2002} and magnetic properties\cite{Dresselhaus2002}. GICs have potential as highly conductive materials\cite{Dresselhaus2002,Vogel1977,Foley1977,Shioya1986}. GICs with metal chloride or pentafluoride intercalants, such as Antimony pentafluoride (SbF$_5$) and Arsenic pentafluoride (AsF$_5$), received much interest since the 1970s\cite{Dresselhaus2002,Vogel1977,Foley1977,Shioya1986}. E.g., AsF$_5$ GICs has slightly higher $\sigma$ (6.3$\times$10$^{5}$Scm$^{-1}$)\cite{Vogel1977} than bulk Cu\cite{Foley1977,Shioya1986} (5.9$\times$10$^{5}$Scm$^{-1}$)\cite{Vogel1977}, while the graphite in plane $\sigma$ is$\sim$4.5$\times$10$^{4}$Scm$^{-1}$ [\onlinecite{Spain1973}]. The $\sigma$ increase is assigned to injection of carriers from the intercalate layer, with low mobility, to the graphite layers, with high mobility\cite{Dresselhaus2002}.

GICs can be superconducting\cite{Hannay1965} with transition temperatures up to 11.5K for CaC$_6$ GICs at ambient pressure\cite{Weller2005}, and higher with increasing pressure\cite{Csanyi}. GICs are also promising for hydrogen storage, due to a larger interlayer spacing\cite{Deng2004}. GICs are already commercialized in batteries\cite{Enoki}, in particular, in Li-ion batteries since the 1970s\cite{Watanabe1970,Besenhard1976,Nobuatsu1980,Winter1998}. GICs have also been used as negative electrodes (anode during discharge) in Li-ion batteries with the introduction of solid electrolytes\cite{Yazami1999,Yazami1983}.

A number of approaches have been developed over the years for GIC production, starting from solid\cite{Hérold1955}, liquid\cite{Ebert1976} or gaseous reagents\cite{Croft1960}. Intercalation requires a high vapor pressure (\textit{i.e.}$\sim$3-5atm) to enable intercalants to penetrate between the graphite layers\cite{Dresselhaus2002,Croft1960}. The most common production strategies include two-zone vapor transport\cite{Dresselhaus2002,Hérold1955,Zhao2011}, exploiting T differences between graphite and intercalants\cite{Croft1960} and, sometimes, the presence of gases\cite{Croft1960}, \textit{e.g.} Cl$_2$ for intercalation of AlCl$_3$\cite{Dresselhaus2002}. GICs can be produced in single (for binary or ternary GICs) or multiple steps, the latter when direct intercalation is not possible\cite{Falardeau1978}. Hundreds of GICs with donor (alkali, alkali earth metals, lanthanides, metal alloys or ternary compounds, etc.) or acceptor intercalants (\textit{i.e.} halogens, halogen mixtures, metal chlorides, acidic oxides, etc.) have been reported\cite{Dresselhaus2002,Zhao2011}.

The intercalation process increases the graphite interlayer spacing, especially for low stage index GICs\cite{Kwon2011,Lerf1998}. E.g., K, Rb or Cs-GICs have interlayer distance$\sim$0.53-0.59nm, while larger intercalants, such as dimethylsulfoxide, give an interlayer distance$\sim$0.9nm\cite{Lerf1998}, \textit{i.e.} 1.5 to$\sim$3 times larger than the$\sim$0.34nm spacing in pristine graphite. This makes GICs promising starting materials to produce graphene via LPE, even without ultrasonication\cite{Valles2008,Kwon2011,Lerf1998,Ang2009,Catheline2011}. However, although the exfoliation process is often called spontaneous\cite{Valles2008,Catheline2011}, due to the absence of an ultrasonication step, it requires mechanical energy, often provided by stirring\cite{Valles2008,Catheline2011}. To date it is possible to exfoliate GICs with lateral sizes$\sim$20$\mu$,m with Y$_M\sim$90$\%$\cite{Ang2009}, and mobilities of $\sim$tens cm$^2$V$^{-1}$[\onlinecite{Ang2009}].

Note that many GICs tend to oxidize in air\cite{Dresselhaus2002,Vogel1977b}, and require a controlled ambient for their processing\cite{Dresselhaus2002,Vogel1977b}. This, coupled with the additional steps for GIC production, is one of the primary reasons why GICs are not yet extensively used to produce graphene via LPE. However, Ref.[\onlinecite{Khrapach2012}] recently reported FeCl$_3$ intercalated FLGs air-stable for up to one year.
\subsection{\label{SiC}Growth on SiC}
The production of graphite from SiC, Fig.\ref{Figure_1}e, was reported by Acheson as early as 1896 (Ref.\onlinecite{Acheson1896}) for lubricant applications\cite{Acheson1896}. The growth mechanism has been investigated since the 1960s\cite{Badami1962,Van Bommel1975}. Both surfaces (Si(0001)- and C(000-1)-terminated) annealed at high temperature ($>$1000$^{\circ}$C) under ultra-high vacuum (UHV) graphitize due to the evaporation of Si\cite{Forbeaux2000,Charrier2002}. Refs.[\onlinecite{Berger2004,Emtsev2009}] reported the production of graphene films by thermal decomposition of SiC above 1000$^{\circ}$C. Thermal decomposition is not self-limiting\cite{Emtsev2009}, and areas of different thicknesses may exist on the same SiC crystal\cite{Emtsev2009}.

On the Si(0001)-face, graphene grows on a C-rich 6$\sqrt{3}$ $\times$ 6$\sqrt{3}$ R30$^{\circ}$ reconstruction with respect to the SiC surface\cite{Emtsev2008}, called "buffer layer"\cite{Emtsev2008}. This consists of C atoms arranged in a graphene-like honeycomb structure\cite{Emtsev2008}, but without graphene-like electronic properties, because$\sim$30$\%$ are covalently bonded to Si\cite{Emtsev2008,Varchon2007}.

The buffer layer can be decoupled from the Si(0001)-face by hydrogen intercalation\cite{Hass2008,Riedl2009,Goler2013} becoming a quasi-free-standing SLG with typical linear $\pi$ bands\cite{Hass2008}.

Growth of graphene on SiC is referred to as "epitaxial growth"\cite{Epitaxial} even though there is a very large lattice mismatch between SiC (3.073${\AA}$) and graphene (2.46${\AA}$), and carbon rearranges in a hexagonal structure as Si evaporates from the SiC substrate, rather than being deposited on the SiC surface as would happen in a traditional epitaxial growth process. The term "epitaxy" derives from the Greek, the prefix epi means "over" or "upon" and taxis means "order" or "arrangement". In 1928 Royer [\onlinecite{Royer1928}] used the term "epitaxy" referring to the "oriented growth of one substance on the crystal surface of a foreign substance". If the growing crystal and the substrate have the same lattice constants these are lattice matched\cite{Bachmann1995}. The use of "epitaxial" as the adjectival form of epitaxy has been subjected to some criticism already in the sixties, because it is incorrect from the philological point of view\cite{Schneider1963}. Epitactic is the correct form\cite{Schneider1963}. In 1965 epitaxic was recommended by Ref.[\onlinecite{Pashley1965}]. However, the word "epitaxial" is now widely used, and any attempt to change it is unrealistic. We will thus use "epitaxial" as adjectival form of epitaxy. There are two epitaxial processes, depending on the substrate: homo- and hetero-epitaxy. In the case of homoepitaxy the substrate is of the same composition and structure as the growing film, whereas in heteroepitaxy the substrate is of a different composition, and may not be perfectly lattice matched.

It would be desirable to grow graphene on a lattice matched isostructural substrate, in order to minimize defects, like misfit dislocations, as in the case of traditional semiconductors\cite{Pashley1956}. However, with the exception of graphite, where the growth would be referred to as homoepitaxy and is not useful or practical for obvious reasons, there are few substrates that are isostructural and nearly lattice matched with graphene. There are two potential subtrates that might meet the aforementioned requirement, h-BN and hexagonal closed packed (hcp) Co. h-BN has the lowest lattice mismatch$\sim$1.7$\%$. Cobalt metal (hcp at T$<$400$^\circ$C) also has a small lattice mismatch$\sim$2$\%$. There are other hcp metals like Ru, Hf, Ti, Zr but these have much larger lattice mismatch\cite{Jain1996} than that between Co and graphene. Face center cubic metals like Ni, Cu, Pd, Rh, Ag, Au, Pt and Ir have a range of lattice mismatch on the (111) planes. Therefore, from an epitaxial growth perspective, it would be desirable to grow on oriented single crystal Co (see Sect.\ref{Precipitation},\ref{CVD}) as performed by Ref.\onlinecite{Ago2010}. Growth on Co would also require transfer to other non-metallic subtrates, as discussed later. SiC could be ideal, were it not for the fact that the lattice mismatch between graphene and SiC is also very large,$\sim$25$\%$, both for 4H-SiC (Si-face) and 6H-SiC (C-face). Perhaps it is not appropriate to call graphene growth on SiC epitaxial, but this is what numerous papers do. There have been reports of growth of layered materials on highly non-lattice-matched substrates as buffer layers, due to their weak bonding to the underlying substrates\cite{Ueno1997,Koma1999,Jaegermann2000}. In this case the films grow parallel to the substrate because of the anisotropic nature of their chemical bonds. Growth of graphene on SiC might be described in a similar manner\cite{Ueno1997,Koma1999,Jaegermann2000}.

The growth rate of graphene on SiC depends on the specific polar SiC crystal face\cite{De Heer2011,De Heer}. Graphene forms much faster on the C- than on the Si-face\cite{De Heer2011,De Heer}. On the C-face, larger domains ($\sim$200nm) of multilayered, rotationally disordered graphene are produced\cite{Hass2006,Hass2006PRL}. On the Si-face, UHV annealing leads to small domains,$\sim$30-100nm\cite{Hass2006,Hass2006PRL}. The small-grain structure is attributed to morphological changes of the surface during annealing\cite{Emtsev2009}.

Different strategies have been proposed to control the Si sublimation rate. Ref.[\onlinecite{Tromp2009}] used Si vapors to establish thermodynamic equilibrium between SiC and external Si vapor, in order to vary the transition T from the Si-rich (3$\times$3) to the C-rich (6$\sqrt{3}\times$6$\sqrt{3}$R30$^{\circ}$) phase, and final graphene layer. The resulting domains were an order of magnitude larger than those grown under UHV\cite{Hass2008}.

Ref.[\onlinecite{Emtsev2009}] used the "light bulb method" to grow graphene, exploiting a 80-year old process first developed to extend the lifetime of incandescent lightbulb filaments\cite{Fonda1923}. This uses Ar in a furnace at near ambient pressure (1 bar) to reduce the Si sublimation rate. Indeed, in Ar no sublimation is observed until 1500$^{\circ}$C, whereas Si desorption starts at 1150$^{\circ}$C in UHV\cite{Emtsev2009}, thus enhancing surface diffusion, with complete surface restructuring before graphene formation\cite{Emtsev2009}. The resulting films on the Si-face have$\sim$50$\mu$m domains\cite{Emtsev2009}, almost 3 orders of magnitude larger than in UHV annealing\cite{Hass2006,Hass2006PRL}.

Si sublimation can also be controlled by confining SiC in an enclosure (either in vacuum\cite{De Heer2011} or inert gas\cite{De Heer2011}) limiting Si escape, maintaining a high Si vapor pressure. This keeps the process close to thermodynamic equilibrium, resulting in either SLG\cite{De Heer2011} or FLG\cite{De Heer2011} over large (cm scale) areas, both on Si-[\onlinecite{De Heer2011}] and C-faces[\onlinecite{De Heer2011}]. High T annealing of SiC can also give GNRs and GQDs\cite{Sorkin2010,Sprinkle2010}.

To date, graphene grown on the Si-face has a RT mobility up to$\sim$500-2000cm$^2$V$^{-1}$s$^{-1}$[\onlinecite{De Heer2011}], with higher values on the C-face ($\sim$10000-30000cm$^2$V$^{-1}$s$^{-1}$)\cite{Hass2006PRL,De Heer2011,De Heer}. For near-intrinsic samples (8.5$\times$10$^{10}$cm$^{-2}$)\cite{Dawlaty2008} RT mobilities up to$\sim$150000cm$^{2}$V$^{-1}$s$^{-1}$ on the C-face\cite{Tedesco2009} and$\sim$5800cm$^{2}$V$^{-1}$s$^{-1}$ on the Si-face\cite{Tedesco2009} were reported.

Graphene on SiC has the benefit that SiC is an established substrate for high frequency electronics\cite{Davis1991}, light emitting devices\cite{Davis1991}, and radiation hard devices\cite{Davis1991}. Top-gated transistors have been fabricated from graphene on SiC on a wafer scale\cite{Kedzierski2008}. High frequency transistors have also been demonstrated with 100GHz cut-off frequency\cite{Li_Science2010}, higher than state-of-the-art Si transistors of the same gate length\cite{Schwierz2010}. Graphene on SiC has been developed as a novel resistance standard based on the quantum Hall effect (QHE)\cite{Zhang2005,Du2009,Novoselov2005b}.

A drawback for this technology to achieve large scale production equivalent to that in the present Si technology, is the SiC wafers cost ($\sim\$$150-250 for 2"\cite{tanke} at 2011 prices, compared to$\sim\$$5-10 for same size Si) and their smaller size (usually no larger than 4"\cite{tanke}) compared to Si wafers. One approach to reduce substrate costs is to grow thin SiC layers on sapphire, the latter costing less than$\sim\$$10 for 2"\cite{Yole}, and subsequently perform thermal decomposition to yield FLG\cite{McArdle2011}. Thus far, FLGs produced in this way have inferior structural and electronic quality compared to those on bulk SiC. Another approach is to grow SiC on Si\cite{Russo2009}. However SiC on Si is usually cubic\cite{Matsunami1978,Boo1995,Nishino1980,Ikoma1991}, making it challenging to achieve continuous high quality graphene, due to bowing and film cracking as a consequence of high residual stress\cite{Ouerghi2010,OuerghiPRB}. Ref.[\onlinecite{Coletti2011}] grew SLG on 3C-SiC(111) with domains$\sim$100$\mu$m$^2$, by combining atmospheric pressure growth\cite{Emtsev2009} with hydrogen intercalation\cite{Riedl2009}, demonstrating that large area domains can also be grown on 3C-SiC(111).
\subsection{\label{Precipitation}Growth on metals by precipitation}
The first reports of synthetic growth of graphite, i.e. not extracted from mined natural sources, on transition metals date back to the early 1940s\cite{Lipson1942,Biscoe1942}. However, the details of the growth process were not elucidated until the 1970's, when Shelton et al.[\onlinecite{Shelton1974}] identified, via a combination of Auger and low-energy electron diffraction (LEED), SLG formed from carbon precipitation, following high T annealing of Co, Pt, Ni. Graphite can also be obtained from carbon saturated molten Fe during the formation of steel\cite{Winder2006}. In this process, Fe is supersaturated with carbon, and the excess carbon precipitates\cite{Winder2006}. This is usually referred to as "Kish graphite"\cite{Walker1957}, from the German "Kies", used by steel workers to refer to the "mixture of graphite and slag separated from and floating on the surface of molten pig iron or cast iron as it cools"\cite{Derbyshire1975}.

The amount of carbon that can be dissolved in most metals is up to a few atomic percent\cite{Massalski1990}. In order to eliminate the competition between forming a carbide and graphite/graphene growth, the use of non-carbide forming metals, e.g. Cu, Ni, Au, Pt, Ir, is preferred\cite{Li_Science}. Elements like Ti, Ta, Hf, Zr and Si, etc. form thermally stable carbides, as shown by the phase diagram\cite{Okamoto1990,Cadoff1953,Fernandez1995,Okamoto2000,Kaufman1979}, thus are not "ideal" for graphite/graphene growth. Moreover, all have a large ($>$20\%) lattice mismatch with graphene.

Carbon can be deposited on the metal surface by a number of techniques, flash evaporation, physical vapor deposition (PVD), chemical vapor deposition (CVD), spin coating. The carbon source can be a solid\cite{Ruan2011,SunNature}, liquid\cite{Guermoune2011,Miyasaka2011,Miyata2010} or gas\cite{Karu1966}.
In the case of pure carbon, flash evaporation\cite{Peters1984} or PVD\cite{Powell1966}, can be used to deposit carbon directly on the substrate of interest, before diffusion at high T, followed by precipitation of graphite (graphene) upon cooling. When the solid source is a polymer, it can be spun on the metal substrate at RT, followed by high T annealing and growth\cite{SunNature}, as mentioned above.

The growth process on Ni was first investigated in 1974 in Ref.[\onlinecite{Shelton1974}]. They observed SLG on Ni(111) at T$>$1000K by Auger analysis, followed by graphite formation upon cooling. During high T annealing, carbon diffuses into the metal until it reaches the solubility limit. Upon cooling, carbon precipitates forming first graphene, see Fig.\ref{Figure_1}f, then graphite\cite{Shelton1974}. The graphite film thickness depends on the metal, the solubility of carbon in that metal, the T at which the carbon is introduced, the thickness of the metal and the cooling rate.

There has been an effort to try and use inexpensive metals to grow large area (cm scale) graphene, such as Ni\cite{Reina2009,Kim2009,Reina2009NR,Yu2008} and Co\cite{Ramon2011} , while growth on noble metals such as Ir\cite{N'Diaye2006}, Pt\cite{Hamilton1980}, Ru\cite{Himpsel1982,Yoshii2011,Gao2007,Gao2009}, and Pd\cite{Hamilton1980,Little2003}, was performed primarily to study the growth mechanism\cite{Sutter2008,Sutter2009,Loginova2008,McCarty2009,Loginova2009}, and/or obtain samples suitable for fundamental studies, e.g. for scanning tunneling microscopy (STM)\cite{Gao2007,Gao2009,Vazquez2008}, requiring a conductive substrate.

Growth of graphene on Ni\cite{Coraux2008,Reina2009,Kim2009,Reina2009NR,Yu2008}, Co\cite{Ramon2011}, Ru\cite{Yoshii2011}, etc. was also reported by so-called chemical vapor deposition at high temperature, using various hydrocarbon precursors\cite{Reina2009,Kim2009,Reina2009NR,Yu2008,Ramon2011}. However, the CVD process referred to in the aforementioned papers is a misnomer, since graphene is not directly produced on the metal surface by the reaction and deposition of the precursor at the "growth T", but rather grows by carbon segregation from the metal bulk, as a result of carbon supersaturation in the solid, as discussed above\cite{Shelton1974,Karu1966}.

For lattice mismatches between graphene and substrate below 2$\%$, commensurate superstructures, where the resulting symmetry (between graphene and substrate) is a doubling of the unit cell along one axis (\textit{i.e.} 1/2, 0,0), are formed\cite{Coraux2008}. This is the case in Co(0001)\cite{Vaari1997}. Larger mismatches yield incommensurate (i.e. with total loss of symmetry in a particular direction, \textit{i.e.}(0.528,0,0)) Moir\`{e} superstructures, such as in Pt(111)\cite{Land1992}, Ir(111)\cite{Busse2011}, or Ru(0001)\cite{Sutter2008}. E.g., high-T segregation of C on Ru(0001) gives a spread of orientations\cite{Sutter2008}. Also, the graphene/Ru lattice mismatch\cite{Vazquez2008} gives a distribution of tensile and compressive strains\cite{Jiang2009}, thus causing corrugation, with a roughness$\sim$2${\AA}$\cite{Jiang2009}. The Moir\`{e} superstructure could be eliminated by adsorption of oxygen\cite{Zhang2009}, since this weakens the graphene interaction with the substrate\cite{Zhang2009}.

Growth of graphene by precipitation requires careful control of the metal thickness, T, annealing time, cooling rate, and metal microstructure. Recently, Ref.[\onlinecite{Yoshii2011}] reported growth on Ni, Co and Ru on sapphire. Through the suppression of grain boundaries, Ref.[\onlinecite{Yoshii2011}] demonstrated uniform growth on Ru by a surface catalyzed reaction of hydrocarbons, but not on Ni and Co\cite{Yoshii2011}. Both SLG and FLG were observed on Ni and Co, presumably due to the higher solubility of carbon and incorporation kinetics in comparison to Ru at the same T\cite{Yoshii2011}. However, Ref.[\onlinecite{Ago2010}] grew graphene on epitaxial Co on sapphire, achieving SLGs, in contrast to FLGs in Ref.[\onlinecite{Yoshii2011}]. An alternative strategy for SLG growth on high C solubility substrates was proposed by Ref.[\onlinecite{Dai2011}], using a binary alloy (Ni-Mo). The Mo component of the alloy traps all the dissolved excess C atoms, forming molybdenum carbides and suppressing C precipitation\cite{Dai2011}. Graphene was also grown on epitaxial Ru(0001) on sapphire\cite{Sutter2010}.

One of the shortcomings of the growth on metals is that most applications require graphene on an insulating substrate. Ref.[\onlinecite{Peng2011}] suggested that graphene can be grown directly on SiO$_2$ by the precipitation of carbon from a Ni film deposited on the dielectric surface. This process has promise but needs further refinement.
\subsection{\label{CVD}Chemical vapor deposition (CVD)}
CVD is a process widely used to deposit or grow thin films, crystalline or amorphous, from solid, liquid or gaseous precursors of many materials. CVD has been the workhorse for depositing many materials used in semiconductor devices for several decades\cite{Kern1979}.

The type of precursor is usually dictated by what is available, what yields the desired film, and what is cost effective for the specific application. There are many different types of CVD processes: thermal, plasma enhanced (PECVD), cold wall, hot wall, reactive, and many more. Again, the type depends on the available precursors, the material quality, the thickness, and the structure needed, plus it is important to keep in mind that cost is an essential part of selecting a specific process.

The main difference in the CVD equipment for the different precursor types is the gas delivery system\cite{Xu2009}. In the case of solid precursors, the solid can be either vaporized and then transported to the deposition chamber\cite{Xu2009}, or dissolved using an appropriate solvent\cite{Xu2009}, delivered to a vaporizer\cite{Xu2009}, and then transported to the deposition chamber\cite{Xu2009}. The transport of the precursor can also be aided by a carrier gas\cite{Xu2009}. Depending on the desired deposition T, precursor reactivity, or desired growth rate, it may be necessary to introduce an external energy source to aid precursor decomposition.

One of the most common and inexpensive production methods is PECVD. The creation of plasma of the reacting gaseous precursors allows deposition at lower T with respect to thermal CVD. However, since plasma can damage the growing material, one needs to design the equipment and select process regimes that minimize this damage. The details of the growth process are usually complex, and in many cases not all of the reactions are well understood. There are many different ways to perform plasma assisted CVD and it is not the objective of this review to cover all of them (see Ref.[\onlinecite{Meyyappan2003}] for an overview). It is however important to match the equipment design with the material one is trying to deposit and the precursor chemistry. Graphene should be simpler than multi-component systems, since it is a single element material. As with many other materials, graphene growth can be performed using a wide variety of precursors, liquids, gases, solids, growth chamber designs, thermal-CVD or PECVD, over a wide range of chamber pressures and substrate T. In the next sections we will describe CVD of graphene on metals and dielectrics.
\subsubsection{\label{Thermal CVD }Thermal CVD on metals }
In 1966 Karu and Beer[\onlinecite{Karu1966}] used Ni exposed to methane at T=900$^\circ$C to form thin graphite, to be used as sample support for electron microscopy. In 1971, Perdereau and Rhead\cite{Perdereau1971} observed the formation of FLG via evaporation of C from a graphite rod\cite{Perdereau1971}. In 1984 Kholin et al.[\onlinecite{Kholin1984}] performed what may be the first CVD graphene growth on a metal surface, Ir, to study the catalytic and thermionic properties of Ir in the presence of carbon\cite{Gall2000}. Since then, other groups exposed metals, such as single crystal Ir\cite{Coraux2008,Charrier1994}, to carbon precursors and studied the formation of graphitic films in UHV systems.

The first studies of graphene growth on metals primarily targeted the understanding of the catalytic and thermionic activities of the metal surfaces in the presence of carbon\cite{Moller1987}. After 2004, the focus shifted to the actual growth of graphene. Low pressure chemical vapor deposition (LPCVD) on Ir(111) single crystals using an ethylene precursor was found to yield graphene structurally coherent even over the Ir step edges\cite{Coraux2008}. While Ir can certainly be used to grow graphene by CVD, see Fig.\ref{Figure_1}g, because of its low carbon solubility\cite{Massalski1990}, it is difficult to transfer graphene to other substrates because of its chemical inertness. Ir is also very expensive. Growth on Ni\cite{Reina2009} and Co\cite{Ramon2011,Orofeo2011}, metals compatible with Si processing since they have been used for silicides for over a decade\cite{Gambino1998,Maex1993,Morimoto1995,Lavoie2003,Zhang2003}, and less expensive than Ir, poses a different challenge, \textit{i.e.} FLGs are usually grown\cite{Karu1966,Charrier1994,Reina2009,Kim2009,Reina2009NR,Yu2008,Ramon2011}, and SLGs grow non-uniformly, as described in Sect.\ref{Precipitation}. Therefore, while many papers claim CVD growth at high T on Ni and Co\cite{Karu1966,Charrier1994,Reina2009,Kim2009,Reina2009NR,Yu2008,Ramon2011}, the process is in fact carbon precipitation, not yielding uniform SLG, rather FLGs. The shortcoming of high solubility or expensive and chemically unreactive metals motivated the search for processes and substrates better suited to yield SLG.

The first CVD growth of uniform, large area ($\sim$cm$^2$) graphene on metal was in 2009 by Ref.[\onlinecite{Li_Science}] on polycrystalline Cu foils, exploiting thermal catalytic decomposition of methane and low carbon solubility. This process is almost self-limited, \textit{i.e.} growth is suppressed as soon as the Cu surface is fully covered with graphene, but still with$\sim$5$\%$ BLG and 3LG\cite{Li_Science,Li2009}. Large area graphene growth was enabled principally by the low C solubility in Cu\cite{Lopez2004}, and Cu mild catalytic activity\cite{Li2009b}.

Growth of graphene on Cu by LPCVD was then scaled up in 2010 by Ref.[\onlinecite{Bae2010}], increasing the Cu foil size (30 inches), producing films with mobility ($\eta$)$\sim$7350cm$^2$V$^{-1}$s$^{-1}$ at 6K. Large grain,$\sim$20-500$\mu$m, graphene on Cu with $\eta$ ranging from$\sim$16,400 to$\sim$25,000cm$^{2}$V$^{-1}$s$^{-1}$ at RT after transfer to SiO$_{2}$ was reported in Refs.[\onlinecite{Petrone2012,Li2010}], and from$\sim$27,000 to$\sim$45,000cm$^{2}$V$^{-1}$s$^{-1}$ on h-BN at RT\cite{Petrone2012}.

The current understanding of the growth mechanism is as follows. Carbon atoms, after decomposition from hydrocarbons, nucleate on Cu, and the nuclei grow into large domains\cite{Li2010,Li2011}. The nuclei density is principally a function of T and pressure and, at low precursor pressure, mTorr, and T$>$1000$^{\circ}$C, very large single crystal domains,$\sim$0.5mm are observed\cite{Li2010,Li2011}. However, when the Cu surface is fully covered, the films become polycrystalline, since the nuclei are not registered\cite{Li_Science,Li2010,Li2011},\textit{i.e.} they are mis-oriented or incommensurate with respect to each other, even on the same Cu grain. This could be ascribed to the low Cu-C binding energy\cite{Frese1987}. It would be desirable to have substrates such as Ru, with higher binding energy with C\cite{Frese1987}. However, while Ru is compatible with Si processing\cite{Aoyama1999}, oriented Ru films may be difficult to grow on large (300-450mm) diameter Si wafers, or transferred from other substrates.

There are some difficult issues to deal with when growing graphene on most metal substrates, especially Cu, because of the difference in thermal expansion coefficient between Cu and graphene, of about an order of magnitude\cite{Yoon2011}. The thermal mismatch gives rise to a significant wrinkle density upon cooling\cite{Li2009}. These wrinkles are defective, as determined by Raman spectroscopy\cite{Li2010}, and may also cause significant device degradation through defect scattering, similar to the effect of grain boundaries on mobility in semiconducting materials\cite{Li2010}. These defects however, may not be detrimental for many non-electrically-active applications, such as transparent electrodes. Perhaps one could use cheaper substrates, such as Cu (Cu is cheaper than Ir, Ru, Pt) and use an electrochemical process to remove graphene while reusing Cu, so that the cost is amortized over many growth runs. Because of some unattractive properties (e.g. surface roughening and sublimation) of Cu at the current thermal CVD growth T$>$1000$^\circ$C, the community has been searching for new substrates that take advantage of the self-limited growth process, in addition to dielectrics. Ref.[\onlinecite{Addou2012}] reported growth of SLG on Ni(111) at lower T, 500-600$^\circ$C, using ethylene by UHV CVD, and identified the process as self-limiting, presumably due to the low C solubility in Ni at T$<$650$^\circ$C\cite{Vertman1965}. However, the T range within which graphene can be grown on Ni is very narrow, $~$100$^\circ$C\cite{Addou2012}, and could result in a Ni$_2$C phase\cite{Addou2012}, which can give rise to defects within the Ni crystal. Thus one could surmise that any graphene growing on the surface could be non-uniform across the Ni-Ni$_2$C regions.

Graphene was also grown on Cu by exposing it to liquids or solid hydrocarbons\cite{SunNature,Li2011ACS}. Ref.[\onlinecite{Li2011ACS}] reported growth using benzene in the T range 300-500$^\circ$C.

The process space for SLG-CVD growth is very wide and depends on many factors, from substrate choice, to specific growth conditions, as well as variables not under direct control. It is critical to know the material requirements for specific applications, so that one can tune the growth process/conditions to the application. Growth of graphene on single crystal substrates would be a desired route for improving electronic properties. Following the growth of graphene on Cu, Ago et al. [\onlinecite{Ago2010}] developed a Co deposition process to form highly crystalline Co on c-plane sapphire, where they grew SLG by CVD at high T. However they did not distinguish between face centered cubic (fcc)(111)Co and hcp(0002)Co and did not comment on potential phase transformation issues at T lower than the fcc to hcp phase transition T$\sim$400$^\circ$C. While this approach may seem incompatible with Si processing, and the material cost could be high, it is important to learn how to take advantage of processes that enable growth of higher quality graphene on stable surfaces, not necessarily single crystals.
	
Another question is: can we controllably grow FLGs$?$ Catalytic decomposition of CO on various metals, such as Fe, Cu, Ag, Mo, Cr, Rh, and Pd, was studied by Kehrer and Leidheiser in 1954\cite{Kehrer1954}. They detected graphitic carbon on Fe after exposure to CO for several hours at 550$^\circ$C, but found the other metals to be inactive. The presence of BLG and TLG on Cu\cite{Li_Science} poses the question of the growth process for these isolated regions, since at first one would like to grow uniformly SLG. Growth of controlled Bernal stacked films is not easy, but small regions have been observed\cite{Lee2010NL}. Ref.[\onlinecite{Lee2010NL}] reported homogenous BLG by CVD on Cu. However, it is not clear whether the films are of high enough quality for high performance electronic devices, since Ref.[\onlinecite{Lee2010NL}] did not report D peak Raman mapping, and $\eta\sim$580cm$^2$ V$^{-1}$ s$^{-1}$ at RT. Another approach was proposed by Ref.[\onlinecite{Liu2011}], by increasing the solubility of C in Cu via a solid solution with Ni, forming the binary alloy, Cu-Ni. By controlling Ni percentage, film thickness, solution T, and cooling rate, N was controlled, enabling BLG growth\cite{Liu2011}.
\subsubsection{\label{CVD insulators}CVD on insulators}
Electronic applications require graphene grown, deposited or transferred onto dielectric surfaces. To date, with the exception of graphene grown on SiC by Si evaporation (see Sect.\ref{SiC}), SLG that can satisfy the most area demanding applications, such as flat panel displays, was grown solely on metals. It is unfortunate that SiC substrates are expensive, of limited size, and that SiC cannot be easily grown on Si or other useful substrates for electronic devices. Therefore, it is necessary to develop direct growth on dielectrics, not involving Si evaporation at high T. Growth of high-quality graphene on insulating substrates, such as SiO$_2$, high-k dielectrics, h-BN, etc. would be ideal for electronics. There have been many attempts to grow on Si$_3$N$_4$\cite{Sun2011}, ZrO$_2$\cite{Scott2011}, MgO\cite{Rummeli2010}, SiC\cite{Strupinski2011}, and sapphire\cite{Fanton2011}. However, while graphitic regions are observed at T$<$1000$^\circ$C, none of the processes yield, to date, planar SLG films covering the whole surface\cite{Scott2011,Fanton2011}. Ref.[\onlinecite{Coleman_KS2011}] used a method that involves spraying a solution of sodium
ethoxide in ethanol under Ar atmosphere into the hot zone ($\sim$900$^\circ$C) of a tube furnace, where the sodium ethoxide decomposes,
and deposits on quartz or Si as FLGs. The films on quartz have a Rs$\sim$4.7K$\Omega$/$\square$ and transmittance$\sim$76$\%$. Ref. [\onlinecite{Coleman_KS2011}] used a similar procedure (just a different concentration of sodium ethoxide) to produce graphene nanoplates in large quantity, soluble in liquids. However, the Raman spectra clearly show the presence of very defective flakes\cite{Coleman_KS2011}. Thus far, the best quality was achieved on sapphire\cite{Fanton2011} (3000cm$^2$V$^{-1}$s$^{-1}$ and 10500cm$^2$V$^{-1}$s$^{-1}$ at RT and 2K, respectively). h-BN was also shown to be an effective substrate\cite{Tanaka2003,Song2010,Shi2010,Liu2011b}, with promise for hetero-epitaxial growth of heterostructures (\textit{e.g.} graphene/h-BN)\cite{Tanaka2003,Liu2011b}.
\subsubsection{\label{Plasma}Plasma enhanced CVD}
Reducing the growth T is important for most applications, especially when considering the process for complementary metal-oxide semiconductor (CMOS) devices. The use of plasmas to reduce T during growth/deposition was extensively exploited in the growth of nanotubes and amorphous carbon\cite{Chhowalla2001,Teo2001,Hofmann2005,Boskovic2002,Casiraghi2007,Casiraghi2003,Moseler2005}. Graphene was grown by PECVD using methane at T as low as 500$^\circ$C\cite{Kim2011,Terasawa2012}, but the films had a significant D-band, thus with quality still not equivalent to exfoliated or thermal CVD graphene\cite{Kim2011,Terasawa2012}. Nevertheless, Ref.[\onlinecite{Kim2011}] demonstrates that growth may be carried out at low T, and perhaps the material can be used for applications not having the stringent requirements of the electronics industry. E.g., Ref.[\onlinecite{Kim2011}] used PECVD at T=317$^\circ$C to make TCs with Rs $\sim$2k$\Omega/\Box$ at 78$\%$ transmittance.

Inductively coupled plasma (ICP) CVD was also used to grow graphene on 150mm Si\cite{Lee2010c}, achieving uniform films and good transport properties (\textit{i.e.} $\eta$ up to$\sim$9000cm$^2$ V$^{-1}$ s$^{-1}$). This process is still under development with, as of this writing, insufficient data on the structure of the material.

In 1998 Ref.[\onlinecite{Burden1998}] reported SLG with a curved structure as a byproduct of PECVD of diamond-like carbon. A number of other groups later produced vertical SLGs\cite{Wang2004} and FLGs\cite{Malesevic2008,French2005,French2006,Chuang2007,Chuang2006,Mori2011} by microwave PECVD on several substrates, including non-catalytic, carbide forming substrates, such as SiO$_{2}$. SLGs and FLGs nucleate at the substrate surface, but then continue to grow vertically, perhaps because of the high concentration of carbon radicals\cite{Kim2011}, thus resulting in high growth rate. This material is promising for supercapacitors or other applications, such as field emission, not requiring planar films.
\subsection{\label{MBE}Molecular beam epitaxy}
Molecular beam epitaxy (MBE) is widely used and well suited for the deposition and growth of compound semiconductors, such as III-V, II-VI\cite{Cho1975}. It was used to grow graphitic layers with high purity carbon sources, Fig.\ref{Figure_1}e, on a variety of substrates such as SiC\cite{Al-Temimy2009}, Al$_2$O$_3$\cite{Jerng2011,Liu2012}, Mica\cite{Seifarth,Lippert2011}, SiO$_2$\cite{Seifarth}, Ni\cite{Garcia2010}, Si\cite{Hackley2009}, h-BN\cite{Garcia2012}, MgO\cite{Jerng2012}, ect., in the 400-1100$^\circ$C range. However, these films have a large domain size distribution of defective crystals\cite{Garcia2010}, with lack of layer control\cite{Garcia2010}, because MBE is not a self-limited process relying on the reaction between the deposited species\cite{Cho1975}. Moreover, the reported RT $\eta$ is thus far very low($\sim$1cm$^2$V$^{-1}$s$^{-1}$)\cite{Jerng2011}. Based on the graphene growth mechanism that we have learned over the past few years on metals\cite{Li_Science,Coraux2008,Li2011,Li2009,Li2009b,Kim2009}, specifically Cu\cite{Bae2010,Li_Science,Li2011}, it is unlikely that traditional MBE can be used to make SLG of high enough quality to compete with other processes discussed above. Since MBE relies on atomic beams of elements impinging on the substrate, it is difficult to prevent, say C, from being deposited on areas where graphene has already grown. Therefore, since MBE is a thermal process, the carbon is expected to be deposited in the amorphous or nanocrystalline phase, rather than as graphene. One might however envisage the use of chemical beam epitaxy (CBE)\cite{Tsang1984} to grow graphene in a catalytic mode, taking advantage of the CBE ability to grow or deposit multiple materials, such as dielectrics\cite{LeeAPL2006} or layered materials, on the top of graphene to form heterostructures.
\subsection{\label{ALD}Atomic layer epitaxy}
Atomic layer epitaxy (ALE) has by large not been a successful technique for semiconductor materials as is MBE. Atomic layer deposition (ALD)\cite{Suntola1989} on the other hand has been extensively used over the past ten years to produce thin layers of nano-crystalline binary metal nitrides (e.g. TaN, TiN)\cite{Ritala_Chem1999,Ritala1999}, and high-k gate dielectrics such as HfO$_2$\cite{Kim2004}. The ALD process can be used to grow controllably very thin, less than 1 nm, films\cite{Suntola1989}, but to our knowledge, single atomic layers have not been commonly deposited on large areas.

Large area,$\sim$cm$^2$, graphene was grown by thermal CVD\cite{Bae2010,Li_Science,Li2011} and PECVD\cite{Kim2011,Terasawa2012} using hydrocarbon precursors. A process dealing with a specific precursor and reactant could in principle be used in the ALE mode. However, to date there are no reports, to the best of our knowledge, of ALE-growth of graphene.
\subsection{\label{amorphous}Heat-driven conversion of amorphous carbon and other carbon-based films}
Heat-driven conversion of amorphous carbon (a-C), hydrogenated a-C (a-C:H), tetrahedral a-C (ta-C), hydrogenated (ta-C:H) and nitrogen doped (ta-C:N) ta-C (for a full classification of amorphous carbons see Refs.[\onlinecite{Ferrari2000,Casiraghi2007}]), to graphene could exploit the extensive know-how on amorphous carbon deposition on any kind of substrates (including dielectrics) developed over the past 40 years\cite{chabot}. This process can be done following two main approaches: 1) Annealing after deposition or 2) Annealing during the deposition.

Post-deposition annealing requires vacuum ($<$ 10$^{-4}$mbar)\cite{Ferrari1999,Ilie2000,Kleinsorge2000,Conway2000,Ilie2001} and T depending on the type of amorphous carbon and the presence of other elements such as  nitrogen\cite{Ilie2000,Kleinsorge2000} or hydrogen\cite{Ferrari1999,Ilie2000,Conway2000,Ilie2001}. Ref.[\onlinecite{Ferrari1999}] demonstrated that ta-C transitions from a sp$^3$-rich to a sp$^2$-rich phase at 1100$^\circ$C, with a decrease in electrical resistivity of 7 orders of magnitude from 10$^7$ to 1$\Omega$ cm.  A lower T suffices for a-C:H ($\sim$300$^\circ$C)\cite{Ilie2000} and ta-C:H($\sim$450$^\circ$C)\cite{Ilie2000}. For ta-C:H a drastic reduction of resistivity is observed from 100$^\circ$C (R$\sim$10$^{10}$$\Omega$cm) to 900$^\circ$C (R=10$^{-2}$$\Omega$ cm)\cite{Ilie2001}.

Refs.[\onlinecite{Barreiro,Westenfelder2011}] used a current annealing process for the conversion. However, they did not report the resulting transport properties.

Annealing during deposition allows sp$^3$ to sp$^2$ transition to happen at lower T than post-deposition annealing ($\sim$200$^\circ$C)\cite{Chhowalla2000,Ferrari2000PRB,Kleinsorge2000,Ilie2001}. Ref.[\onlinecite{Chhowalla2000}] reported a reduction of resistivity of$\sim$6 orders of magnitude (R$\sim$10$^8$$\Omega$ cm at RT and R$\sim$10$^2$$\Omega$ cm at$\sim$450$^\circ$C). As in the case of post-processing, the presence of hydrogen (ta-C:H) or nitrogen (ta-C:N) changes the transition T\cite{Kleinsorge2000}. Ref.[\onlinecite{Kleinsorge2000}] demonstrated transition for ta-C:N at$\sim$200$^\circ$C, with a much larger reduction, with respect to ta-C, of resistivity ($\sim$11 orders of magnitude, R$\sim$10$^8$$\Omega$ cm at RT and R$\sim$10$^{-3}$$\Omega$ cm at$\sim$250$^\circ$C, the latter R value comparing well with RGO films\cite{Su2009}). However, unlike post-deposition annealing, annealing during deposition tends to give graphitic domains perpendicular to the substrate\cite{Ilie2001}.

Heat-driven conversion can also be applied to self-assembled monolayers (SAMs), composed of aromatic rings\cite{Turchanin2011}. Ref.[\onlinecite{Turchanin2011}] reported that a sequence of irradiative and thermal treatments cross-links the SAMs and then converts them into nanocrystalline graphene after annealing at 900$^\circ$C. However, the graphene produced via heat-driven conversion of SAM had defects and low mobility ($\sim$0.5cm$^2$V$^{-1}$s$^{-1}$ at RT)\cite{Turchanin2011}. Thus, albeit being simple and cost effective, at the moment the quality of the obtained material is poor, and more effort is needed targeting reduction of structural defects.
\subsection{\label{Chemical}Chemical synthesis}
Graphene can also be chemically synthesized, assembling polycyclic aromatic hydrocarbons (PAHs)\cite{Wu2007,Cai2010,Yan2010}, through surface-mediated reactions, Fig.\ref{Figure_1}i.

Two approaches can be used. The first exploits a dendritic precursor transformed in graphene by cyclodehydrogenation and planarization\cite{Zhi2008}. This produces small domains, called nanographene (NG)\cite{Zhi2008}. The second relies on PAH pyrolysis\cite{Cai2010,Dossel2011}. Other benzene-based precursors, such as poly-dispersed hyperbranched polyphenylene\cite{Inoue2000}, give larger flakes\cite{Zhi2008}.

PAHs can also be exploited to achieve atomically precise GNRs\cite{Cai2010,Dossel2011} and GQD\cite{Yan2010}. The first were synthesized through oxidative cyclodehydrogenation with FeCl$_3$\cite{Zhi2008}. The presence of alkyl chains makes these GNRs soluble\cite{Dossel2011}. The formation of GQDs is more complex, and starts from the synthesis of dendrimers\cite{Yan2010}. More details are in Ref.[\onlinecite{Yan2010}].

The formation of graphene, GNRs and GQDs is mediated by a metal surface acting as catalyst for the thermal reactions occurring at high T\cite{Cai2010}.

Ref.[\onlinecite{Yan2010}] reported GNRs with well-defined band gap and/or GQDs with tuneable absorption, and tested these in solar cells. Chemical synthesis may ultimately allow a degree of control truly at the atomic level, while still retaining scalability to large areas. However, NGs tend to form insoluble aggregates due to strong inter-flakes attraction\cite{Wu2007,Yan2010,Zhi2008}. An approach to solubilize conjugated systems is lateral attachment of flexible side chains\cite{Yan2010}. This has been successful in solubilizing small NGs\cite{Wu2007}, while failing for larger ones\cite{Wu2007}, because the inter-graphene attraction overtakes the intermolecular forces\cite{Israelachvili}. An alternative consists in covalent attachment of multiple 1,3,5-trialkyl-substituted phenyl moieties to NG edges to achieve highly soluble large GQDs\cite{Yan2010}.
\subsection{\label{ribbons}Nano-ribbons and quantum dots}
\begin{figure}
\centerline{\includegraphics[width=90mm]{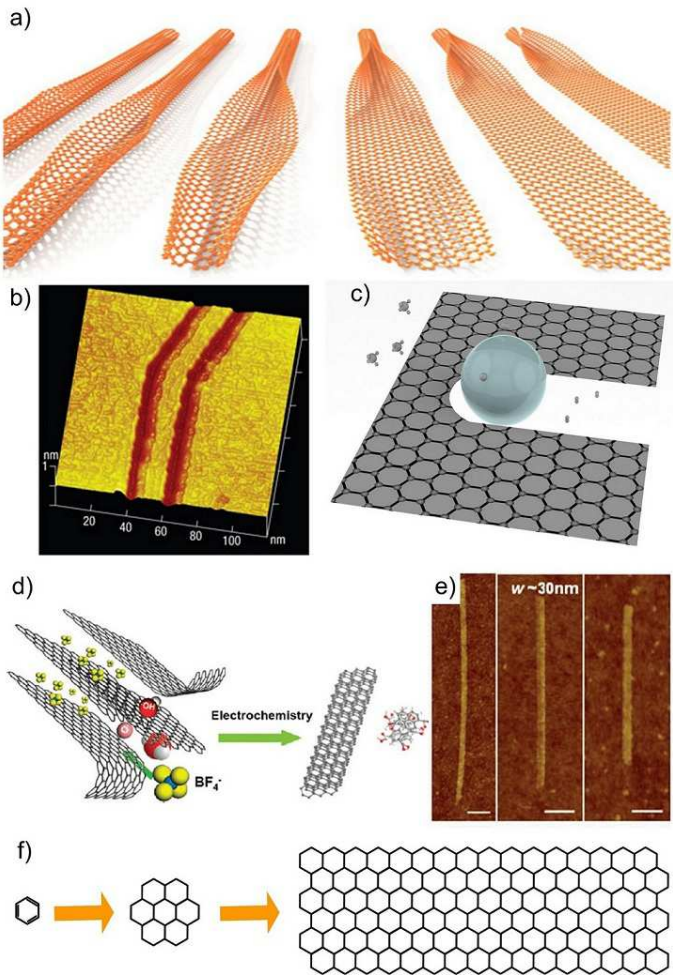}}
\caption{\label{Figure_7} Top-down fabrication of GNRs via (a) unzipping of nanotubes (adapted from Ref.[\onlinecite{Sinitskii2010}]), (b) STM lithography \cite{Tapaszto2008}, (c) catalytic hydrogenation, using thermally activated Ni nanoparticles, (d) exfoliation of chemically modified (adapted from Ref.[\onlinecite{Lu2009}]) and (e) expanded graphite (adapted from Ref.[\onlinecite{Li2008}]). (f) Bottom-up fabrication of GNRs}
\end{figure}

Refs.[\onlinecite{Han2007,Chen2007}] prepared GNRs by combining e-beam lithography and oxygen plasma etching. GNRs down to$\sim$20nm were reported, with band gap$\sim$30meV, then used in FETs with I$_{ON}/I_{OFF}$ up to 10$^3$ at low T ($<$5K) and$\sim$10 at RT. Ref.[\onlinecite{Ponomarenko2008}] reported much smaller GNRs, with minimum width$\sim$1nm and gap$\sim$500meV produced by e-beam lithography and repeated over-etching. Sub-10nm GNRs with bandgap up to 400meV were produced via a chemical route\cite{Li2008}, consisting in the dispersion of expanded graphite in liquid phase followed by sonication. Used as channels in FETs, they achieved  I$_{ON}/I_{OFF}$ up to 10$^7$ at RT\cite{Li2008}. A solution-based oxidative process was also reported\cite{Kosynkin2009}, producing GNRs by lengthwise cutting and unraveling single (SWNTs) and multiwall carbon nanotubes\cite{Sinitskii2010}, Fig.\ref{Figure_7}a . As result of the oxidative process, such GNRs show poor conductivity ($\sim$35S/cm) and low $\eta$ (0.5-3cm$^2$ V$^{-1}$ s$^{-1}$) at RT\cite{Sinitskii2009}.

Patterning of SLG into sub-10nm GNRs with predetermined crystallographic orientation was achieved by STM lithography\cite{Tapaszto2008}, Fig.\ref{Figure_7}b, by applying a bias, higher than for imaging, between STM tip and substrate, while moving the tip at constant velocity.

GNRs can also be formed without cutting. Ref.[\onlinecite{Singh2009}] demonstrated that spatial selective hydrogenation can be used to create graphene "nanoroads", i.e. conductive paths of graphene surrounded by fully hydrogenated areas. Ref.[\onlinecite{Lee2011}] fabricated encapsulated$\sim$35nm GNRs by depositing a polymer mask via scanning probe lithography, followed by chemical isolation of the underlying GNR by fluorinating the uncovered graphene. These GNRs retained the carrier mobility of non-patterned graphene. Also, the fluorination is reversible, enabling write-erase-rewrite. GNRs down to 12nm were produced by local thermal reduction of GO by scanning probe\cite{Wei2010}.

Sub-10nm GNRs were fabricated via catalytic hydrogenation, using thermally activated Ni nanoparticles as "knife"\cite{Ci2008,Campos2009} (Fig.\ref{Figure_7}c). This allows cutting along specific crystallographic directions, therefore the production of GNRs with well-defined edges.

GNRs were also made via LPE of GICs\cite{Lu2009} (Fig.\ref{Figure_7}d) and expanded graphite\cite{Li2008} (Fig.\ref{Figure_7}e). Growth on controlled facets on SiC resulted in 40nm GNRs and the integration of 10,000 top-gated device on a single SiC chip\cite{Spinkle2010}.

Chemical synthesis (Fig.\ref{Figure_7}f) seems to be the most promising route towards well-defined GNRs\cite{Cai2008}. Atomically precise GNRs were produced by surface-assisted coupling of molecular precursors into linear polyphenylenes and subsequent cyclo-de-hydrogenation\cite{Cai2008}. GNRs up to 40nm in length and soluble in organic solvents, such as toluene, dichloromethane and tetrahydrofuran, were synthesized\cite{Dossel2011} from polyphenylene precursors, having a non-rigid kinked backbone, to introduce higher solubility than linear poly(para-phenylene)\cite{Rehahn1989}.

Another route to GNRs is the so-called nanowire lithography\cite{Fasoli2009}, consisting in the use of nanowires as masks for anisotropic dry etching. GNRs smaller than the wire itself can be fabricated via multiple etching\cite{Fasoli2009}. Also, the wire, consisting of a crystalline core surrounded by a SiOx shell, can be used as self-aligned gate\cite{Kulmala2011}.

Arrays of aligned GNRs have been produced by growing graphene by CVD on nanostructured Cu foils and subsequently transferring on flat Si/SiO$_2$ substrates\cite{Pan2011}. The Cu structuring results in controlled wrinkling on the transferred material\cite{Pan2011}, which allows production of aligned GNRs by plasma etching\cite{Pan2011}.

Besides their semiconducting properties, GNRs show other interesting properties, such, e.g., magnetoelectric effects\cite{Zhang2009b}. Also, half-metallic states can be induced in zigzag GNRs subjected to an electric field\cite{Son2006}, chemically modified zigzag GNRs or edge-functionalized armchair GNRs\cite{Cervantes2008}. Half-metals, with metallic behavior for electrons with one spin orientation and insulating for the opposite, may enable current spin-polarization\cite{Son2006}.

Another approach to tune the bandgap of graphene relies in the production of GQDs\cite{Yan2010,Pan2010,Zhu2011chem,Shen2011,Li2011adv,Lu2011Nat,Liu2011jacs}. These have different electronic and optical properties with respect to pristine graphene\cite{Geim2007,Bonaccorso2010} due to quantum confinement and edge effects.
Graphene oxide quantum dots (GOQDs) have been produced via hydrothermal\cite{Pan2010} and solvothermal\cite{Zhu2011chem} methods, with lateral sizes$\sim$10nm and in the 5-25nm range, respectively. Another route to produce GOQDs exploited the hydrazine hydrate reduction of small GO sheets with their surface passivated by oligomeric polyethylene glycol (PEG)\cite{Shen2011}. These GOQDs show blue PL under 365nm excitation, while green fluorescence was observed with 980nm excitation\cite{Shen2011}.
GOQDs were also produced by electrochemical oxidation of a graphene electrode in phosphate buffer solution\cite{Li2011adv}. These have heights between 1 and 2nm and lateral size of 3-5nm\cite{Li2011adv}. A bottom-up approach was used by Ref.[\onlinecite{Lu2011Nat}] to produce GQDs by metal-catalysed opening of C$_{60}$. The fragmentation of the embedded C$_{60}$ molecules at T$\sim$550$^\circ$C produced carbon clusters that underwent diffusion and aggregation to form GQDs.

As reported in Sect.\ref{Chemical}, GQDs can also be chemically synthesized, assembling PAHs\cite{Wu2007,Yan2010}, through surface-mediated reactions.
Ref.[\onlinecite{Liu2011jacs}] exploited chemical synthesis to produce GOQDs by using hexa-peri-hexabenzocoronene (HBC) as precursor. The as-prepared GOQDs with ordered morphology were obtained by pyrolysis and exfoliation of large PAHs\cite{Liu2011jacs}. The HBC powder was first pyrolyzed at a high T, then the artificial graphite was oxidized and exfoliated, followed by reduction with hydrazine\cite{Liu2011jacs}. The obtained GOQDs had diameter$\sim$60nm and thickness$\sim$2–3nm, with broad PL\cite{Liu2011jacs}.
\section{\label{processing}Graphene processing after production}
\subsection{\label{Transfer}Transfer, placement and shaping}
The placement of graphene on arbitrary substrates is key for applications and characterization. The ideal approach would be to directly grow graphene where required. However, as discussed above, we are still far from this goal, especially in the case of non-metallic substrates. Alternatively, a transfer procedure is necessary. This also allows the assembly of novel devices and heterostructures, with different stacked 2d crystals.
\subsubsection{\label{Suspended}Suspended graphene}
Graphene membranes are extremely sensitive to small electrical signals\cite{Ponomarenko2008}, forces or masses\cite{Chen2009mass} due to their extremely low mass and large surface-to-volume ratio, and are ideal for nanoelectromechanical systems (NEMS). Graphene membranes have also been used as support for TEM imaging\cite{Meyer2008Nature} and as biosensors\cite{Schneider2010,NairAPL97}. Nanopores in SLGs membranes have been exploited for single-molecule Deoxyribonucleic acid (DNA) translocation\cite{Schneider2010}, paving the way to devices for genomic screening, in particular DNA sequencing\cite{Dekker2012,Well2012}. Thanks to its atomic thickness graphene enables to detect variations between two bases in DNA\cite{Schneider2010}, unlike conventional Si$_3$N$_4$ nanopores\cite{Albrecht2011}.

Freestanding graphene membranes were first reported in Ref.\onlinecite{Ferrari2006}. Graphene samples were deposited by MC onto Si+SiO$_2$ substrates, then a grid was fabricated on them by lithography and metal deposition. Si was then etched by tetramethylammonium hydroxide, leaving a metal cantilever with suspended graphene. This process was originally developed to fabricate suspended SWNT\cite{Meyer2005}. Ref.[\onlinecite{Meyer2007}] used the same approach to fabricate graphene membranes and study them by TEM. Ref.[\onlinecite{Bunch2007}] prepared mechanical resonators from SLG and FLG by mechanically exfoliating graphite over trenches in SiO$_2$. Ref.[\onlinecite{Meyer2008}] transferred graphene exfoliated either on SiO$_2$ or polymer on TEM grids by first adhering the grid, and subsequently etching the substrate. Ref.[\onlinecite{Booth2008}] fabricated graphene membranes up to 100$\mu$m in diameter by exfoliating graphite on a polymer and subsequently fabricating a metal scaffolds on it by e-beam lithography and metal evaporation. The polymer was then dissolved leaving suspended graphene membranes\cite{Booth2008}. A similar technique was used in Ref.[\onlinecite{NairS320}] to produce suspended sample to study graphene's optical transmission. Ref.[\onlinecite{Du2008}] suspended graphene by contacting it via lithography and subsequently etching a trench underneath. This approach allowed to achieve ballistic transport at low T ($\sim$4K)\cite{Du2008} and the highest $\eta$ to date (10$^6$cm$^2$ V$^{-1}$ s$^{-1}$)\cite{BoloSSC146,Elias2011}. Suspending graphene drastically reduces electron scattering, allowing the observation of fractional QHE (FQHE)\cite{Du2009,Bolotin2009}.
\begin{figure}
\centerline{\includegraphics[width=90mm]{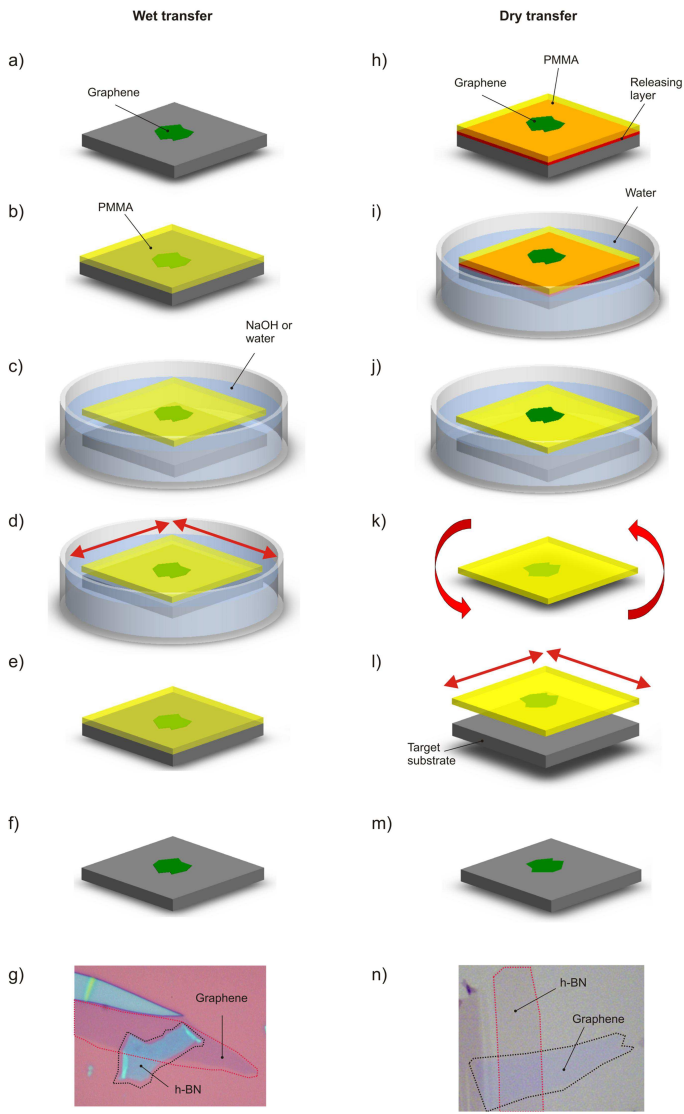}}
\caption{\label{Figure_8} \textbf{Wet transfer}: (a) MC-SLG on Si/SiO$_2$. (b) A PMMA film is deposited by spin coating. (c) The PMMA film is detached either via NaOH etching or water intercalation. Graphene adheres to the polymer and is removed from the Si/SiO$_{2}$ substrate. (d) The PMMA+graphene film is attached to the target substrate. By sliding the PMMA+graphene film with respect to the substrate, the selected flake can be aligned with features, such as electrodes, cavities, etc. (e) Once the sample has dried, (f) PMMA is dissolved by acetone releasing the SLG on the target substrate. (g) A flake deposited onto a BN crystal by wet transfer. \textbf{Dry transfer}: (h) graphene is exfoliated onto a water-dissoluble polymer (such as PVA) covered with PMMA. (i) The sample is left to float in a water bath so that the water soluble layer is dissolved from the side. (j) SLG on top of the stack never touches the water. (k) The polymer+graphene film is attached to an holder and flipped over. (l) By means of a manipulator, the flake of choice is placed in the desired position on top of the desired substrate, then the film is pressed on the target substrate. (m) PMMA is dissolved leaving SLG in the desired position. (n) SLG deposited onto BN by dry transfer [adapted from Ref.\onlinecite{Dean2010}]}
\end{figure}
\subsubsection{\label{Transfer layers}Transfer of individual layers}
Several transfer processes have been developed so far and can be classified either as "wet" or "dry". The first includes all procedures where graphene is in contact, at some stage of the process, with a liquid. In the second, one face of graphene is protected from contacting any liquid, while the other is typically in contact with a polymer, eventually dissolved by solvents.
\subsubsection{\label{Wet Transfer}Wet Transfer of exfoliated flakes}
In 2004 Ref.[\onlinecite{Meitl2004}] placed SWNTs onto arbitrary substrates by transfer printing using poly(dimethysiloxane) (PDMS) stamps. Ref.[\onlinecite{Jiao2008}] reported transfer of various nanostructures (such as SWNTs, ZnO nanowires, gold nanosheets and polystyrene nanospheres) by means of a poly(methyl methacrylate) (PMMA)-mediated process. In 2008 Ref.[\onlinecite{Reina2008}] adapted this process to transfer MC-graphene on various target substrates. The process is based on a PMMA sacrificial layer spin-coated on graphene. The polymer-coated sample is then immersed in a NaOH solution, which partially etches the SiO$_{2}$ surface releasing the polymer. Graphene sticks to the polymer, and can thus be transferred. PMMA is eventually dissolved by acetone, thus releasing graphene.

Ref.[\onlinecite{Bonaccorso2010}] reported the deterministic placement of graphene by exploiting a thin layer of water between the PMMA/graphene foil and the substrate. Ref.[\onlinecite{Dekker2010}] reported transfer of nanostructures (including graphene) embedded in a hydrophobic polymer. Also in this case, intercalation of water at the polymer-substrate interface was used to detach the polymer/nanostructures film, then moved on a target substrate\cite{Dean2010}.

Fig.\ref{Figure_8} shows the steps of a typical wet transfer process, together with an optical picture of a graphene flake transferred on a BN crystal. Deterministic transfer allows fabrication of devices by placing flakes of choice onto pre-patterned electrodes.

\begin{figure}
\centerline{\includegraphics[width=90mm]{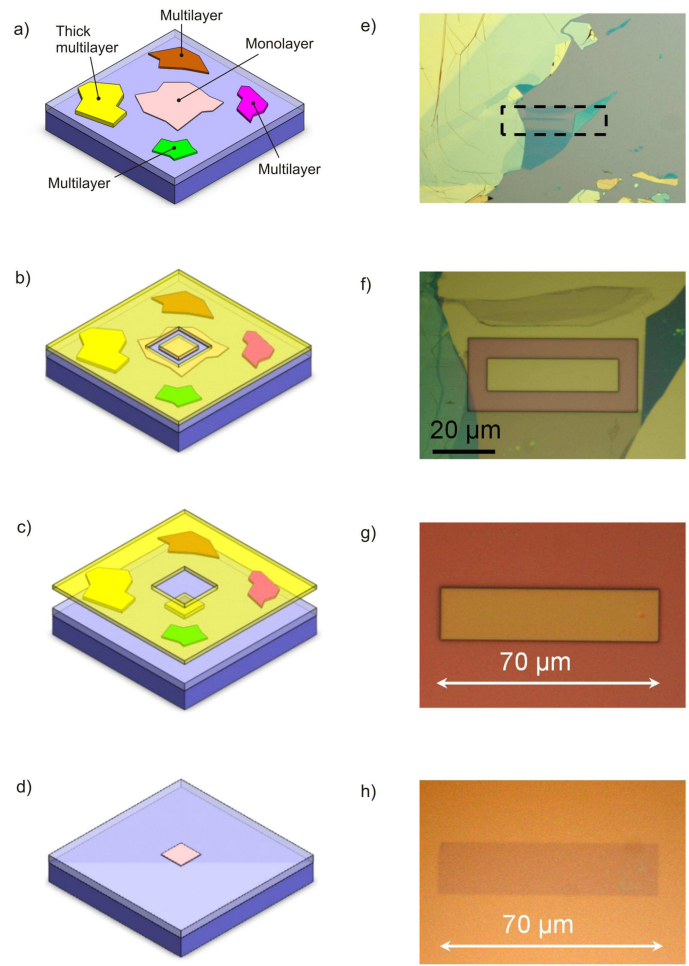}}
\caption{\label{Figure_9} Isolation and shaping of graphene. (a) MC-SLGs are always surrounded by thicker flakes. (b) A PMMA layer is deposited and patterned via e-beam lithography to define a contour of the desired shape. The uncovered flakes are removed by plasma etching. (c) The polymer film is removed by immersion in DI water. The lithographically defined graphene+PMMA island stays on the substrate. (d) The remaining PMMA is dissolved leaving an isolated and shaped graphene layer. (e) SLG surrounded by thicker flakes. The same SLG flake (f) after patterning and etching and (g) after PMMA removal. (h) Final isolated shaped SLG}
\end{figure}

PMMA is a positive resist widely used for high resolution e-beam lithography\cite{Yasin2001}. By patterning PMMA, it is also possible to remove unwanted graphitic material surrounding MC-SLGs, while shaping and isolating the flakes of interest, Fig.\ref{Figure_9}.
\subsubsection{\label{Dry Transfer}Dry Transfer of exfoliated flakes}
In order to fabricate heterostructures with clean interfaces (\textit{i.e.} without trapped adsorbates), dry transfer methods have been developed. Ref.[\onlinecite{Dean2010}] reported a mechanical transfer based on stacking two polymer layers, the bottom being water dissolvable and the top being PMMA. Graphene was exfoliated onto this polymer stack and the sample floated on the surface of deionized (DI) water, resulting in the detachment of the PMMA+graphene film from the substrate. The upper graphene face was not in contact with water, thus minimizing contamination. The polymer+graphene film was then collected and the alignment achieved using a micromanipulator. Ref.[\onlinecite{Mayorov2011a}] used a similar technique to encapsulate graphene between two h-BN layers, while Ref.[\onlinecite{ZomeAPL99}] reported an alternative technique based on a glass/tape/copolymer stack.

Fig.\ref{Figure_8} summarizes both wet and dry transfer processes. However, Ref.[\onlinecite{Haigh2012}] reported that even dry transfer may not result in perfectly clean interfaces, as some adsorbates may get trapped.
\subsubsection{\label{transfer from metals}Transfer of graphene grown on metals}
In 2009 Ref.[\onlinecite{Reina2009}] first reported the transfer of SLG and FLG grown by precipitation on Ni, by depositing a PMMA sacrificial layer and subsequently etching the underlying Ni by aqueous HCl solution. Ref.[\onlinecite{Li2009b}] transferred films grown by CVD on Cu, etched by iron nitrite. Ref.[\onlinecite{Kim2009}] introduced etching by aqueous FeCl$_3$ in order to remove Ni without producing hydrogen bubbles, which may damage graphene when acid etching is used. Ref.[\onlinecite{Kim2009}] also reported a technique where Polydimethylsiloxane (PDMS) "stamps" are attached directly to the graphene surface. Ni is then chemically etched by FeCl$_3$ leaving graphene attached to the PDMS. Graphene is then transferred to SiO$_2$ by pressing and peeling the PDMS stamp. Ref.[\onlinecite{Bae2010}] introduced roll-to-roll (R2R) transfer of graphene grown by CVD on Cu foils as large as 30x30 inches$^2$, guided through a series of rolls: a thermal release tape was attached to the Cu+graphene foil, and then an etchant removed Cu. The tape+graphene film was then attached to a (flexible) target substrate (e.g. PET) and the supporting tape removed by heating, thus releasing graphene onto the target substrate. In order to avoid Fe contamination caused by FeCl$_3$ etching, ammonium persulfate [(NH$_4$)$_2$S$_2$O$_8$] was used\cite{Aleman2010}. To avoid mechanically defects caused by R2R transfer, a hot pressing process was developed\cite{Kang2012a}: similar to a R2R process, the Cu+graphene foil is first attached to thermal release tape and then Cu is chemically etched. The tape+graphene foil is then placed on the target substrate and they are inserted between two hot metal plates with controlled T and pressure. This results in the detachment of the adhesive tape with very low frictional stress, therefore less defects, than a R2R process\cite{Kang2012a}.
\subsubsection{\label{Di-electrophoresis}Dielectrophoresis}
Electrophoresis is a technique used for separating particles according to their size and electrical charge\cite{Johns1995}. An uniform electric current is passed through a medium that contains the particles\cite{Johns1995}. These travel through a medium at a different rate, depending on their electrical charge and size. Separation occurs based on these differences\cite{Johns1995}. Di-electrophoresis (DEP) is the migration of uncharged particles towards the position of maximum field strength in a non-uniform electric field\cite{Pohl1951}. The force in DEP depends on the electrical properties of the particle and surrounding fluid, the particle geometry, and electric field frequency\cite{Johns1995}. Particles move towards regions of high electric field strength (positive DEP) if their polarizability is greater than the suspending medium\cite{Johns1995}, whereas they move in the opposite direction (negative DEP) if the polarizability is lower\cite{Johns1995}. This allows fields of a particular frequency to manipulate particles\cite{Johns1995}, at the same time assembling them on pre-defined location\cite{Johns1995}.

In 2003 Ref.[\onlinecite{Krupke2003}] reported large area deposition of SWNTs between electrode pairs by DEP\cite{Krupke2003}. Subsequently, DEP was used for the separation of metallic (m-SWNTs) and semiconducting (s-SWNTs)\cite{Krupke_Science2003}, exploiting their dielectric constants difference, resulting in opposite movement of m-SWNTs and s-SWNTs\cite{Krupke_Science2003}. These SWNT processes were then adapted for graphene. Refs. [\onlinecite{Burg2009,Kang2009}] used DEP for the manipulation of GO soot, and single- and few-layer GO flakes. In 2009 Ref.[\onlinecite{Vijayaraghavan2009}] placed individual FLGs between pre-patterned electrodes via DEP. Once trapped, the higher polarizability of graphene compared to the surrounding medium\cite{Vijayaraghavan2009} limited the deposition to one flake per device\cite{Hong2008,Vijayaraghavan2009}. This self-limiting nature is one of the advantages of this method, together with the direct assembly of individual flakes at predetermined locations. On the other hand, DEP does not allow deposition/coating on large areas.
\subsubsection{\label{inks}Placement of dispersions and inks}
Dispersions or inks can be used in a variety of placement methods, including vacuum filtration, spin and spray coating, ink-jet printing and various R2R processes. The latter are most attractive because of their manufacturing characteristics, with transfer speeds in excess of 5ms$^{-1}$ currently used in a variety of applications\cite{Cheng2009}. R2R consists in processing and printing a rapidly moving substrate\cite{Cheng2009,Tracton2006}. Generally, a flexible substrate (e.g. paper, textile, polymer) is unrolled from a source roller, coated (\textit{i.e.} without patterning) or printed (\textit{i.e.} with patterning), with one or more evaporated materials (\textit{e.g.} dielectrics) or liquid inks (e.g. inks containing polymers or nanoparticles), simultaneously or in sequence, and treated/cured while the substrate continuously moves along the coating/printing roller, before being rolled up again, or cut into individual pieces/devices. Unlike assembly style "pick and place" strategies, the continuous fabrication process makes R2R a cheap technology\cite{Tracton2006}, ideal for high throughput coating, printing and packaging. R2R is a focus of research in plastic electronics, because of its high throughout, and low cost of ownership compared to other approaches (e.g. conventional vacuum deposition and lithography pattern technologies) with similar resolution\cite{Sheats2002,Choi2008}. A standard R2R process may include evaporation, plasma etching, spray or rod-coating, gravure, flexographic, screen or inkjet printing and laser patterning\cite{Cheng2009}. In many R2R processes, e.g., rod-coating or flexographic printing, solution processing of the ink or material (e.g. polymer, nanoparticles) is required, especially when they cannot\cite{Tracton2006} be evaporated at low T\cite{Cheng2009,Tracton2006,Wong2009}.

Rod-coating employs a wire-wound bar, also known as Mayer bar (invented by Charles W. Mayer who also founded the Mayer Coating Machines Company in 1905 in Rochester, USA)\cite{Tracton2006}. This is a stainless steel rod wound with a tight wire spiral, also made of stainless steel. During coating, this creates a thin ($\sim$tens $\mu$ms) ink layer on a substrate\cite{Sheats2002}. Spray coating forms aerosols, resulting in uniform thin ($\sim$$\mu$m) films on a substrate\cite{Tracton2006}. Screen printing, on the other hand, uses a plate or 'screen' containing the pattern to be printed on the substrate\cite{Tracton2006}. The screen is then placed onto the target substrate, while the ink is spread across the screen using a blade, thus transferring the pattern\cite{Tracton2006}. Flexo- and gravure\cite{Wong2009} printing also use a plate to transfer images onto target substrates. Flexo uses a relief plate, usually made of flexible polymeric material, where the raised sections are coated with ink, then transferred onto the substrate by contact printing\cite{Tracton2006}. Gravure, derived from the Italian word "intaglio" that means engraved or cut-in, uses an engraved metallic plate, consisting of dots representing pixels\cite{Tracton2006,Mitzi2009}. The physical volume of the engraved dots defines the amount of ink stored in them\cite{Mitzi2009}, thus can be used to create gray-scale patterns/images\cite{Mitzi2009}. In general, different viscosities are preferred for different R2R techniques, ranging from 1mPas-10,000mPas or above\cite{Tracton2006,Mitzi2009}. Rod- or spray-coating form uniform films, that may be used for larger scale devices ($>$several cm), the fabrication of TCs, or devices such as batteries or supercapacitors. Screen ($\sim$50-100$\mu$m resolution\cite{Tracton2006}), flexographic ($\sim$40$\mu$m resolution\cite{Tracton2006}) and gravure (15$\mu$m resolution\cite{Tracton2006}) printing can be used to print different materials with specific patterns for flexible electronics\cite{Mitzi2009}. For resolution down to$\sim$50$\mu$m, inkjet printing offers a mask-less, inexpensive and scalable low-T process\cite{Noh2007}. The resolution can be significantly enhanced ($<$500nm) by pre-patterning the substrates\cite{Noh2007}, so that the functionalized patterns can act as barriers for the deposited droplets\cite{Noh2007}. The volume can be reduced to atto-liters/drop by pyroelectrodynamic printing\cite{Ferraro2010}. The process is based on the control of local pyroelectric forces, activated by scanning a hot tip or a laser beam over a functionalized substrate (\textit{e.g.} lithium niobate\cite{Ferraro2010}), drawing droplets from the reservoir, depositing them on the substrate underside\cite{Ferraro2010}.

All the above techniques can be applied to graphene inks/dispersions. Large scale placement of LPE graphene can be achieved via vacuum filtration\cite{Hernandez2008}, spin\cite{Eda2008} and dip coating\cite{Wang2007}, Langmuir-Blodgett\cite{Li2008W} and spray coating\cite{Blake2008}. Amongst the R2R techniques, rod-coating has been demonstrated to fabricate TCs. Inkjet printing was also reported\cite{Torrisi2012}. The advantages of inkjet printing include greater ease of selective deposition and high concentration for partially soluble compounds\cite{Calvert2001}. Ref.[\onlinecite{Torrisi2012}] reported an inkjet printed graphene TFT with $\eta$ up to$\sim$95cm$^2$V$^{-1}$s$^{-1}$, and 80$\%$ transmittance. Inkjet printing of GO was also demonstrated\cite{Jo2011,Luechinger2008,Wang2010,Yu2011}. To minimize clustering of the flakes at the nozzle edge, these should be smaller than 1/50 of the nozzle diameter[\onlinecite{Torrisi2012}].
\subsection{\label{Cleaning}Contamination and cleaning}
Cleaning is a critical part of semiconductor device processing\cite{Kern}. It is usually performed after patterning and etch processes leave residues\cite{Kern}. Wet chemical etches are also performed to remove damage from surfaces\cite{Kern}. Most applications require graphene on a dielectric surface. When graphene is grown directly on a dielectric as in the case of graphene on SiC [see Sect.\ref{SiC}] or when graphene or GO is deposited on the dielectric substrate directly[see Sect.\ref{inks}], cleaning is required only after patterning and etch processes as devices are fabricated.

Because every atom is a surface atom, graphene is very sensitive to contaminants left by production, transfer or fabrication processes. In order to remove them, several methods have been developed.
\subsubsection{\label{Cleaning MC}Cleaning of graphene produced by MC}
The amount of contamination can be assessed optically\cite{Bruna2009}: organic contamination arising from the diffusion of tape glue used in MC changes the contrast\cite{Bruna2009}. TEM and scanning probe\cite{Ishigami2007,Stolyarova2007} microscopies (e.g. AFM, STM), Raman\cite{Casiraghi2007APL,Cheng2011}, together with transport measurements\cite{Cheng2011}, are other viable techniques to detect contaminants on graphene films or flakes. Ref.[\onlinecite{Ishigami2007}] cleaned MC samples from resist residuals by thermal annealing (at 400$^\circ$C, in Ar/H$_2$), assessing the quality of the cleaning process via scanning probe techniques. Ref.[\onlinecite{Stolyarova2007}] introduced thermal annealing (at 280$^\circ$C) in ultra-high vacuum ($<$1.5$\times$10$^{-10}$Torr), to remove resist residues and other contaminants. Ref.[\onlinecite{Moser2007}] cleaned graphene by using high current ($\sim$10$^8$A/cm$^2$). This allows removal of contamination in-situ, and is particularly useful when graphene devices are measured in a cryostat\cite{Moser2007}. Chemical cleaning by chloroform was reported in Ref.[\onlinecite{Cheng2011}]. Mechanical cleaning by scanning the graphene surface with an AFM tip in contact mode was also reported\cite{Goossens2012}.
\subsubsection{\label{Cleaning transfer_MC}Cleaning after transfer}
Cleaning is particularly important when transferring flakes, as the processes typically involves sacrificial layers, to be chemically dissolved, see Sect.\ref{Dry Transfer}. Thermal annealing in H$_2$/Ar is normally used\cite{Mayorov2011a,Haigh2012,Fuhrer2011}.

In graphene transfer from metals to dielectric surfaces, organic materials such as PMMA or Perylene-3,4,9,10-tetracarboxylic dianhydride (PTCDA) are typically used as the carrier material, with subsequent chemical removal, e.g. by acetone\cite{Li_Science,Li2009b}. Great care must be taken to ensure that PMMA or PTCDA are completely removed. Refs.[\onlinecite{Pirkle2011,Chan2012}] detected by XPS (X-ray photoelectron spectroscopy) the presence of residue on the surface of graphene grown on Cu and transferred onto SiO$_2$. The C1s spectrum was found broader than that of graphite and the original graphene on Cu. The broadening was associated with the presence of the residue. Upon annealing in high vacuum (1$\times$10$^{-9}$ mbar) at T$\sim$300$^\circ$C, the C1s width decreased to a value close to the original graphene on Cu\cite{Pirkle2011,Chan2012}. The use of thermal release tape\cite{Bae2010} (i.e. a tape adhesive at RT, but that peels off when heated), instead of PMMA or PTCDA, is more problematic, since tape residues can contaminate the transferred graphene\cite{Bae2010}. There is some anecdotal evidence that the presence of residue has a beneficial effect on the nucleation of ALD dielectrics, such as Al$_2$O$_3$\cite{Lee2009}. However, this approach to prepare the graphene surface for ALD is not ideal, since the residues have uncontrolled chemical nature and are not uniform. Ref.[\onlinecite{Liang2011}] developed a modified RCA transfer method, combining an effective metal cleaning process with control of the hydrophilicity of the target substrates. RCA stands for Radio Corporation of America, the company that first developed a set of wafer cleaning steps in the semiconductor industry\cite{Kern}. Ref.[\onlinecite{Liang2011}] demonstrated that RCA offers a better control both on contamination and crack formation with respect to the aforementioned approaches\cite{Bae2010,Li_Science,Li2009b}.
\subsubsection{\label{Cleaning LPE}Removal of solvents/surfactants in LPE graphene}
For graphene and GO produced via LPE, the cleaning, removal of solvents and/or surfactants, mainly depends on the target applications. For composites (both for mechanical\cite{Stankovich2006} and photonic\cite{Hasan2009,SunZ2010,Hasan2010} applications) the presence of surfactants does not compromise the mechanical and optical properties, thus their removal is not needed, and is in fact essential to avoid agglomeration\cite{Stankovich2006,Hasan2009,SunZ2010,Hasan2010}. Different is the situation when the applications require high conductivity ($>$10$^4$Scm$^{-1}$), \textit{i.e.} TCFs. In this case, the presence of solvents/surfactants compromises the interflake connections, decreasing the electrical performance. The solvents and the deposition strategy (see Sect.\ref{inks}) used for the TCFs production mostly determines the cleaning procedure. In the case of TCFs produced by vacuum filtration (\textit{i.e.} on a cellulose filter membrane) of surfactant-assisted aqueous dispersions, the as-deposited graphene or RGO films are first rinsed with water to wash out the surfactants\cite{Lotya2009,Green2009} and then transferred from the membrane to the target substrate. The membrane is then usually dissolved in acetone and methanol\cite{Wu2004}. For freestanding films, the deposited flakes are peeled off from the membrane\cite{Lotya2009}. The films are then annealed at T$>$250$^\circ$C in Ar/N$_2$\cite{Lotya2009} or air\cite{Green2009}. The latter process could help remove residual surfactant molecules\cite{Lotya2009,Green2009}. However, there is not a "fixed" T for the removal of the solvents/surfactants, and the different conditions/requirements are essentially ruled by the boiling/melting points of each solvent/surfactant.
\section{\textbf{Inorganic layered compounds and hybrid structures}}
\subsection{\label{LM}2d crystals}
There are several layered materials, whose bulk properties were studied already in the sixties\cite{Wilson1969}, which retain their stability down to monolayers, and whose properties are complementary to those of graphene.

Transition metal oxides (TMOs) and transition metal dichalcogenides (TMDs) have a layered structure\cite{Wilson1969}. Atoms within each layer are held together by covalent bonds, while van der Waals interactions hold the layers together\cite{Wilson1969}. LMs include a large number of systems with interesting  properties\cite{Wilson1969}. E.g., NiTe$_2$ and VSe$_2$ are semi-metals\cite{Wilson1969}, WS$_2$, WSe$_2$, MoS$_2$, MoSe$_2$, MoTe$_2$, TaS$_2$, RhTe$_2$, PdTe$_2$ are semiconductors\cite{Wilson1969}, h-BN, and HfS$_2$ are insulators, NbS$_2$, NbSe$_2$, NbTe$_2$, and TaSe$_2$ are superconductors\cite{Wilson1969}; Bi$_2$Se$_3$, Bi$_2$Te$_3$ show thermoelectric properties\cite{Wilson1969} and may be topological insulators\cite{Kane2005}.

Due to the weak bonding of the stacked layers\cite{Wilson1969}, LMs were mainly used as solid lubricants because of their tribological properties\cite{Prasad1997}. In addition, LMs were also used as thermoelectric materials\cite{Mishra1997}, in batteries\cite{Poizot2000}, electrochemical\cite{Abruna1982} and photovoltaics (PV) cells\cite{Djemail1981}, light emitting diodes\cite{Frey2003}, as ion exchangers\cite{Clement1978}, photocatalysts\cite{Kudo1999}, etc.

Similar to graphite and graphene, the LM properties are a function of N. E.g., bulk MoS$_2$ has an indirect band gap\cite{Gmelin1995}, while a monolayer has a direct band gap\cite{Splendiani2010,Mak2010}, that could be exploited for optoelectronics\cite{sundaram}.

We now highlight the most promising routes for production of 2d crystals and hybrids.
\subsubsection{\label{MC LM}Micromechanical cleavage}
As with graphene, individual layers can be made by MC\cite{Novoselov2005}. MC can involve a single crystal\cite{Schultz1994}, or a single grain\cite{Schultz1994}, in the case of polycrystalline materials\cite{Suryanarayana1995}. The local scale dynamics of the fracture process is complex\cite{Schultz1994} and depends on the crystal structure\cite{Schultz1994}. To date the lateral size of 2d crystals produced via MC is$\sim$10$\mu$m in h-BN\cite{Pacile2008}, limited by the average crystal size of the starting material\cite{Pacile2008}. Similar size flakes ($\sim$10$\mu$m) were also achieved via MC of MoS$_2$, WS$_2$ and NbSe$_2$\cite{Benameur2011}.

As in the case of MC of graphite, MC of LMs is not industrially scalable, and MC-flakes are mostly suited for fundamental studies and proof of concept devices.
\subsubsection{\label{LA LM}Laser ablation}
Ref.[\onlinecite{Castellanos2012}] used laser pulses to ablate TMDs (MoS$_2$) down to a single-layer. Ref.[\onlinecite{Castellanos2012}] generated single layer-MoS$_2$ [1L-MoS$_2$] in arbitrary shapes and patterns with feature sizes down to 200nm, with electronic and optical properties comparable to MC-1L-MoS$_2$\cite{Benameur2011}. Indeed, Ref.[\onlinecite{Castellanos2012}] reported similar PL emission between MC-1L-MoS$_2$ and laser thinned 1L-MoS$_2$, and $\eta$ up to 0.49cm$^2$ V$^{-1}$ s$^{-1}$ and 0.85cm$^2$ V$^{-1}$ s$^{-1}$ for laser thinned and MC-1L-MoS$_2$, respectively.
\subsubsection{\label{LPE LM}Liquid Phase Exfoliation}
LPE can exfoliate LMs in organic solvents\cite{Coleman2011,Han2008,Lin2010b,Warner2010,Zhi2009,Cunningham2012} and aqueous solutions\cite{Smith2011,Lin2011}, with\cite{Smith2011}, or without\cite{Lin2011} surfactants, or their mixtures\cite{Zhou2011}. The exfoliated sheets can then be stabilized against re-aggregation either by interaction with solvent\cite{Coleman2011}, or through electrostatic repulsion due to the adsorption of surfactant molecules\cite{BonaDGU,Smith2011}. In the case of solvent stabilization, Ref.[\onlinecite{Coleman2011}] showed that the best solvents are those having surface tension matching the surface energy of the target LM. The dispersions can then produce inks with a variety of properties\cite{Torrisi2012}, or be used for processing in thin films and/or composites\cite{Coleman2011}.

Research is still at an early stage, and LPE must be extended to a wider range of materials. To date, exfoliated TMDs both in organic solvents\cite{Coleman2011} and surfactant-aqueous solutions\cite{Smith2011} tend to exist as multilayers\cite{Coleman2011,Smith2011}. Yields can be defined for LPE of LMs in the same way as done for graphene, see Sect.\ref{LPE}. Ref.[\onlinecite{ONeill2012}] reported Y$_W$=40$\%$ for MoS$_2$ dispersions. To the best of our knowledge, thus far no data exist for Y$_M$ and Y$_{WM}$ in LPE TMOs and TMDs. In order to determine Y$_W$ a library of $\alpha$ for all LMs must be produced. To date, $\alpha$ values were suggested only for few LMs, \textit{i.e.} MoS$_2$ (3400mLmg$^{-1}$ m$^{-1}$), WS$_2$ (2756mLmg$^{-1}$ m$^{-1}$) and BN (2367mLmg$^{-1}$ m$^{-1}$)\cite{Coleman2011}. However, as in the case of graphene, there is uncertainly in the determination of $\alpha$. E.g., Ref.[\onlinecite{Coleman2011}] reported $\alpha\sim$3400mLmg$^{-1}$ m$^{-1}$ for MoS$_2$, while Ref.[\onlinecite{ONeill2012}] used 1020mL mg$^{-1}$ m$^{-1}$.

As in the case of graphene, the development of a sorting strategy in centrifugal fields, both in lateral dimensions and N, will be essential.

Refs.[\onlinecite{Frey2003,Liu1984,Joensen1986,Zahurak1987,Eda2011}] used intercalation of alkali metals with TMD crystals (\textit{i.e.} MoS$_2$ and WS$_2$) to increase the inter-layer distance and facilitate exfoliation.

\subsubsection{\label{Synthesis LM}Synthesis by thin film techniques}
A number of thin film processes can be brought to bear on the growth of 2d crystals. These range from PVD (e.g. sputtering), evaporation, vapor phase epitaxy, liquid phase epitaxy, chemical vapor epitaxy, MBE, ALE, and many more, including plasma assisted processes. The selection of the growth process depends on the material properties needed and the application. Each material has its own challenges and we do not aim here to describe each one individually. However, we note that, other than controlling the thickness and orientation of the films, their composition and stoichiometry is of utmost importance, having large influence on transport. As a result, great care must be taken in controlling the point defect concentration. Low growth T ($\sim$300$^\circ$C) techniques are usually better suited in controlling the defects arising from vacancies, since the vapor pressure of the chalcogenide elements decreases exponentially with T\cite{Li2012}. However, low T techniques tend to give higher extended defect densities because of the lower atomic mobility\cite{Kizilyalli1999}. Therefore, the growth technique must be selected to match the application very carefully.

To date, WS$_2$ films have been deposited by magnetron sputtering from both WS\cite{Buck1991} and WS$_2$ targets\cite{Rai1997}, sulfurization of W\cite{Matthäus1996} or WO$_2$ films\cite{Genut1992}, ion beam mixing\cite{Hirano1985}, etc. The preferred production process for tribological applications is magnetron sputtering\cite{Ellmer2008}, because of its lower T than thermally activated deposition methods\cite{Ellmer2008}. Magnetron sputtering is also well suited for large area deposition ($\sim$m$^2$)\cite{Szczyrbowski1999}. CVD was used to grow h-BN\cite{Nagashima1995}, and is now being developed to grow TMDs\cite{Zhang2012}. If single layers of the binary films are desired, then ALD or, more appropriately, ALE might be better suited.
\subsubsection{\label{NR LM}2d crystals nanoribbons}
Nanoribbons based on 2d crystals can also be made, with tuneable electrical and magnetic properties\cite{Li2004,Zheng2010}. MoS$_2$ nanoribbons were produced via electrochemical/chemical synthesis\cite{Li2004}, while zigzag few- and single-layer BN nanoribbons were produced via unzipping multiwall BN nanotubes through plasma etching\cite{Zheng2010}.
\subsection{\label{Hybrid}Graphene and other 2d crystals hybrids}
Technological progress is determined, to a great extent, by developments in material science. The most surprising breakthroughs are attained when a new type of material, or new combinations of known materials, with different dimensionality and functionality, are created. Well-known examples are the transition from three-dimensional (3d) semiconducting structures based on Ge and Si to 2d semiconducting heterostructures, nowadays the leading platform for microelectronics. Ultimately, the limits and boundaries of certain applications are given by the very properties of the materials naturally available to us. Thus, the band-gap of Si dictates the voltages used in computers, and the Young's modulus of steel determines the size of the construction beams. Heterostructures based on 2d crystals will decouple the performance of particular devices from the properties of naturally available materials. 2d crystals have a number of exciting properties, often unique and very different from those of their three-dimensional (3d) counterparts. However, it is the combinations of such 2d crystals in 3d stacks that offer truly unlimited opportunities in designing the functionalities of such heterostructures. One can combine conductive, insulating, probably superconducting and magnetic 2d materials in one stack with atomic precision, fine-tuning the performance of the resulting material. Furthermore, the functionality of such stacks is "embedded" in the design of such heterostructures.
\begin{figure}
\centerline{\includegraphics[width=90mm]{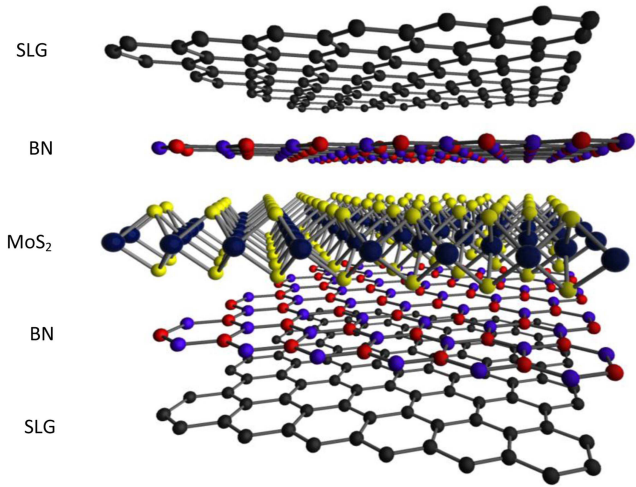}}
\caption{\label{Figure_10} Schematic hybrid superstructure (SLG/BN/MoS$_2$/BN/SLG).}
\end{figure}

Heterostructures have already played a crucial role in technology, giving us semiconductor lasers and high-mobility field effect transistors (FET). However, thus far the choice of materials has been limited to those which can be grown (typically by MBE) one on top of another, thus limiting the types of structures which can be prepared. Instead, 2d crystals of very different nature can be combined in one stack with atomic precision, offering unprecedented control on the properties and functionalities of the resulting 2d-based heterostructures. 2d materials with very different properties can be combined in one 3d structure, producing novel, multi-functional materials. Most importantly, the functionality of such heterostructures will not simply be given by the combined properties of the individual layers. Interactions and transport between the layers allow one to go beyond simple incremental improvements in performance and create a truly "quantum leap" in functionality. By carefully choosing and arranging the individual components one can tune the parameters, creating materials with tailored properties, or "materials on demand". Following this novel approach, part of the functionality is brought to the level of the design of the material itself.

Inorganic LMs can be exploited for the realization of heterostructures with graphene, to modulate/change the electronic properties, thus creating "materials on demand": hybrid superstructures, with properties not existing in nature\cite{Mayorov2011a,Ponomarenko2011,Ramasubramaniam2011,Britnell2012}, tailored for novel applications. E.g., superstructures like those in Fig.\ref{Figure_10} (SLG/BN/MoS$_2$/BN/SLG) can be used for tunnel devices, such as diodes, FETs, and light emitting devices, or for energy application, such as PV cells.

To date, 3 methods can be envisaged for the production of atomically thin heterostructures: (I) growth by CVD\cite{Tanaka2003}; (II) layer by layer stacking via mechanical transfer\cite{Mayorov2011a,Dean2010,Ponomarenko2009} and (III) layer by layer deposition of chemically exfoliated 2d crystals. However, as the field develops, other techniques will emerge.

Field effect vertical tunnelling transistors based on graphene heterostructures with atomically thin BN acting as a tunnel barrier, were reported\cite{Britnell2012}. The device operation relies on the voltage tunability of the tunnel density of states in graphene and of the effective height of the tunnel barrier adjacent to the graphene electrode\cite{Britnell2012}. Ref.[\onlinecite{Georgiou2012}] used Ws$_2$ as an atomically thin barrier, allowing switching between tunneling and thermionic transport, with much better transistor characteristics with respect to the MoS${_2}$ analogue\cite{Britnell2012}, thus allowing much higher ON/OFF
ratios ($\sim$10$^6$). A "barristor", a graphene-silicon hybrid three-terminal device that mimics a triode operation, was developed by Ref.[\onlinecite{kim2012science}].
The electrostatically gated graphene/silicon interface induce a tunable Schottky barrier that controls charge transport across a vertically stacked structure\cite{kim2012science,Britnell2012}.
\subsubsection{\label{CVD_Hybrid}CVD growth of heterostructures}
CVD is suitable for mass production of heterostructures\cite{Liu2011NL,Shi2012}, though it requires the largest investment and effort in terms of identifying the precursors, CVD system design and process development. There are several indications that it is indeed feasible\cite{Liu2011NL,Shi2012}. h-BN was shown to be effective as a substrate for CVD  as well as exfoliated graphene\cite{Kim2011b}.
\subsubsection{\label{MT_Hybrid}Mechanical transfer}
Transfer individual 2d crystals into heterostructures enabled the observation of several interesting effects, including FQHE\cite{Dean2010}, ballistic transport\cite{Mayorov2011a} and metal-insulator transition in graphene\cite{Ponomarenko2011}. An advantage of "dry" mechanical transfer is the possibility to control/modify each layer as it is being deposited, including chemical modifications, at any stage of the transfer procedure. Also, any atomic layer in the multilayer stack can be individually contacted, offering precise control on the properties of the stack (in principle giving a material with individual contacts to every atomic plane). Furthermore, one can apply local strains to individual layers. These can significantly modify their electronic structure\cite{Pereira2009,Mucha2011,Mayorov2011}. Also important is the control of the layer relative orientation, which may affect electronic properties of the stack in certain intervals of the energy spectrum\cite{Luican2011}.
\subsubsection{\label{Inks_Hybrid}Heterostructures from dispersions and inks}
Large-scale placement of LPE samples for the production of heterostructures can be achieved exploiting the techniques reported in Sect.\ref{inks}, tuning the properties of the dispersion/ink accordingly. E.g., surface modifications by self-assembled monolayers enable targeted large-scale deposition. High uniformity and well defined structures on flexible substrates can also be obtained. DEP can be used to control the placement of individual crystals between pre-patterned electrodes. Inkjet printing allows to mix and print layers of different materials and is a quick and effective way of mass-production. Solvothermal synthesis, \textit{i.e.} synthesis in a autoclave using non-aqueous precursors\cite{Chen2007Solvo}, of MoS$_2$ deposited on RGO sheets suspended in dispersion was recently reported\cite{Li2011JACS}.
\section{\label{Conclusion}Outlook and future challenges}
The successful introduction of graphene and/or other 2d materials in products depends not only on the identification of the right products for new and current applications, but also on the ability to produce any of the materials in large quantities at a reasonable cost. The progress in developing new materials processes over the past few years has been impressive, especially given the broad materials requirements, from single crystal graphene to graphene flakes. The suitability of any given process depends on the application. Nanoelectronics more than likely has the most demanding requirements, i.e. low defect density single crystals. Other applications, such as biosensors, may require defective graphene, while printable electronics can tolerate lower quality, e.g. lower mobility, graphene. Chemical vapour deposition techniques are emerging as ideal processes for large area graphene films for touch screens and other large display applications, while graphene derived from SiC single crystals may be better suited for resistor standards. Many issues still remain to be addressed in the growth of graphene by CVD to improve the electrical and optical characteristics, including mechanical distortions, stable doping, and the development of reliable low cost transfer techniques. While transfer techniques can be developed to transfer graphene layers onto insulating substrates, it is desirables to grow graphene directly on dielectric surfaces for many device applications and progress is being made in achieving films on hexagonal boron nitride as well as SiO$_2$. However, a lot more effort is required to achieve large area uniform high quality graphene films on dielectrics. In the case of graphene on SiC, among other issues related to uniformity, crystal size could be a major cost impediment for large scale production. Liquid phase exfoliation is appealing for the production of inks, thin films and composites, and future research is needed to control on-demand the number of layers, flake thickness and lateral size, as well as rheological properties. Synthetic graphenes are the most promising for the production of atomically precise nanoribbons and quantum dots to overcome the lack of band gap necessary for many device applications. A controlled dopant distribution is also needed, and techniques such as surgace functionalization using self-assembled acceptor/donor molecules or assembling pre-doped molecules are being studied.

The layered nature of graphite makes its integration with other layered materials a natural way to create heterostructures. Layered materials have been around for a long time and studied and developed mostly for their tribological properties. Now they are being considered as new interlayer dielectrics for heterostructures that have potential for new electronic devices with exotic properties. Because of their potential for new devices, there will be a host of new processes that will need to be developed in order to grow or deposit high quality large area monolayer films integrated with graphene with controlled thickness and transport properties.
\section{\label{Acknowledgments}Acknowledgments}
We acknowledge M. Bruna, A. Colli, R. Sundaram, R. Weatherup, K. S. Novoselov, for useful discussions and funding from ERC NANOPOTS, EPSRC EP/GO30480/1 and EP/F00897X/1, EU RODIN and GENIUS, the Newton Trust, the Royal Academy of Engineering, King's College, Cambride, the NRI program and a Royal Society Wolfson Research Merit Award.

\end{document}